\documentclass[acmsmall]{acmart}

\setcopyright{acmlicensed}
\copyrightyear{2026}
\acmYear{2026}
\acmDOI{XXXXXXX.XXXXXXX}

\usepackage{colortbl}
\usepackage{pifont}
\usepackage{ifthen}
\usepackage{multirow}
\usepackage{booktabs}
\usepackage{adjustbox}
\usepackage[inline]{enumitem}
\usepackage[ruled,linesnumbered]{algorithm2e}
\usepackage{tcolorbox}
\usepackage{booktabs}
\usepackage{tabularx}
\newcolumntype{L}{>{\raggedright\arraybackslash}X}%
\newcolumntype{R}{>{\raggedleft\arraybackslash}X}%
\newcolumntype{C}{>{\centering\arraybackslash}X}%
\usepackage{listings}
\usepackage{amsfonts}
\usepackage{relsize}
\usepackage{threeparttable}
\usepackage{mathtools}
\usepackage{amsmath}
\usepackage{tikz}
\usepackage{fontawesome5}
\usepackage{subcaption}
\usepackage{xspace}
\usepackage{makecell}
\usepackage{pifont}
\usepackage{soul}
\usepackage{pgf}
\usepackage{svg}
\usepackage{hhline}
\usepackage{seqsplit}

\usetikzlibrary{shapes,positioning}
\makeatletter
\let\old@lstKV@SwitchCases\lstKV@SwitchCases
\def\lstKV@SwitchCases#1#2#3{}
\makeatother
\usepackage{lstlinebgrd}
\makeatletter
\let\lstKV@SwitchCases\old@lstKV@SwitchCases

\lst@Key{numbers}{none}{%
    \def\lst@PlaceNumber{\lst@linebgrd}%
    \lstKV@SwitchCases{#1}%
    {none:\\%
        left:\def\lst@PlaceNumber{\llap{\normalfont
                \lst@numberstyle{\thelstnumber}\kern\lst@numbersep}\lst@linebgrd}\\%
        right:\def\lst@PlaceNumber{\rlap{\normalfont
                \kern\linewidth \kern\lst@numbersep
                \lst@numberstyle{\thelstnumber}}\lst@linebgrd}%
    }{\PackageError{Listings}{Numbers #1 unknown}\@ehc}}
\makeatother

\ifthenelse{\isundefined{\editmode}}{
\newcommand{\editnote}[3]{%
}

}{
\newcommand{\editnote}[3]{\xspace\colorbox{#1}{\sffamily \smaller \textcolor{white}{~\faCommentDots{}~#2~}}\textcolor{#1}{~#3}\xspace}

}

\definecolor{mygreen}{HTML}{0000ff}
\definecolor{darkorange}{HTML}{FE6741}

\definecolor{catppuccin-latte-peach}{HTML}{fe640b}
\definecolor{catppuccin-latte-lavender}{HTML}{7287fd}
\definecolor{crimson}{HTML}{DC143C}

\definecolor{nord0}{HTML}{2E3440}
\definecolor{nord1}{HTML}{3B4252}
\definecolor{nord2}{HTML}{434C5E}
\definecolor{nord3}{HTML}{4C566A}
\definecolor{nord4}{HTML}{D8DEE9}
\definecolor{nord5}{HTML}{E5E9F0}
\definecolor{nord6}{HTML}{ECEFF4}
\definecolor{nord7}{HTML}{8FBCBB}
\definecolor{nord8}{HTML}{88C0D0}
\definecolor{nord9}{HTML}{81A1C1}
\definecolor{nord10}{HTML}{5E81AC}
\definecolor{nord11}{HTML}{BF616A}
\definecolor{nord12}{HTML}{D08770}
\definecolor{nord13}{HTML}{EBCB8B}
\definecolor{nord14}{HTML}{A3BE8C}
\definecolor{nord15}{HTML}{B48EAD}

\definecolor{bestgreen}{HTML}{92C2BC}
\definecolor{secondgreen}{HTML}{B9DCD8}
\definecolor{thirdgreen}{HTML}{DFF0F0}

\newcommand{\code}[1]{\texttt{\seqsplit{#1}}}

\definecolor{boxbg}{RGB}{240, 240 ,240}

\newcounter{rq}

\definecolor{codegreen}{rgb}{0.0510,0.4039,0.0510}
\definecolor{codepurple}{rgb}{0.0510,0.4039,0.0510}
\definecolor{codeblue}{rgb}{0,0.4,0.8}
\definecolor{codered}{rgb}{0.6,0,0}
\definecolor{backcolour}{rgb}{1,1,1}
\definecolor{highlightpink}{rgb}{1.0, 0.8, 0.8}
\definecolor{highlightgreen}{rgb}{0.7255, 0.878, 0.647}

\lstdefinestyle{mystyle}{
    backgroundcolor=\color{backcolour},   
    commentstyle=\color{codegreen},
    keywordstyle=\color{codeblue}\bfseries,
    stringstyle=\color{codepurple},
    moredelim=**[is][\color{codegreen}]{+}{+},
    moredelim=**[is][\color{codered}]{-}{-},
    basicstyle=\ttfamily\footnotesize,
    breakatwhitespace=false,         
    breaklines=true,                 
    captionpos=b,                    
    keepspaces=true,                 
    numbers=left,                    
    numbersep=5pt,                  
    showspaces=false,                
    showstringspaces=false,
    showtabs=false,                  
    tabsize=2,
    frame=lines,
    belowcaptionskip=-3mm, % Adjust this value to decrease space
}

\newcommand{\eg}[1]{(e.g., #1)}
\newcommand{\ie}[1]{(i.e., #1)}

\begin{document}

\title{Finding the Pivots to Repair Broken Source--Test Evolution Using LLMs}

\author{Xiao Chen}
\email{xchenfu@cse.ust.hk}
\orcid{0000-0003-4248-5977}
\affiliation{%
  \institution{The Hong Kong University of Science and Technology}
  \city{Hong Kong SAR}
  \country{China}
}

\author{Songqiang Chen}
\email{i9s.chen@connect.ust.hk}
\orcid{0000-0002-1220-8728}
\affiliation{%
  \institution{The Hong Kong University of Science and Technology}
  \city{Hong Kong SAR}
  \country{China}
}

\author{Jialun Cao}
\email{jcaoap@cse.ust.hk}
\orcid{0000-0003-4892-6294}
\affiliation{%
  \institution{The Hong Kong University of Science and Technology}
  \city{Hong Kong SAR}
  \country{China}
}

\author{Yun Peng}
\email{yunp@fudan.edu.cn}
\orcid{0000-0003-1936-5598}
\affiliation{%
  \institution{Fudan University}
  \city{Shanghai}
  \country{China}
}

\author{Chun Yong Chong}
\email{Chong.chunyong@monash.edu}
\orcid{0000-0003-1164-0049}
\affiliation{%
  \institution{Monash University Malaysia}
  \city{Subang Jaya}
  \country{Malaysia}
}

\author{Shing-Chi Cheung}
\authornote{Corresponding author.}
\email{scc@cse.ust.hk}
\orcid{0000-0002-3508-7172}
\affiliation{%
  \institution{The Hong Kong University of Science and Technology}
  \city{Hong Kong SAR}
  \country{China}
}

\begin{abstract}
  
Repairing obsolete tests in response to source code evolution is critical for test suite reliability and software quality assurance. 
While large language models (LLMs) have emerged as promising tools for automating many software development tasks, the effective supply of context and the orchestration of repair pipelines in this source--test co-evolution scenario remain poorly understood. 
Moreover, although reasoning is found helpful for LLMs on some complex coding tasks, it remains unclear which reasoning steps actually contribute to effective test repairs and how the insights learned from LLMs' behaviors benefit coding agents. 
In this paper, we present a systematic empirical study of test repair for source--test co-evolution empowered by LLMs, investigating the effective setups and typical LLM behaviors. 
To support this study, we construct a new benchmark of 500 real-world source--test co-evolution cases from commits submitted between May 2024 and April 2026 across 15 highly-starred open-source Java projects. 
Our experimental results reveal several key insights for effective LLM-based test repair: the complementary benefits of combining test failure symptoms, source code changes, and supportive code context, as well as the effectiveness of self-repair rounds.
We further derive a common reasoning action taxonomy from LLMs' reasoning traces, which proves instrumental in guiding LLMs to repair tests more effectively while at a lower financial cost. 
Applying our findings to the Codex agent further improves its repair success rate on challenging cases by 66.7\%.

\end{abstract}

%% CCS concepts generated via https://dl.acm.org/ccs



\maketitle

\section{Introduction}\label{sec:1}

% 1. Background
In the development of software systems, test cases verify the correctness of code functionalities, serving as a standard for software quality assurance~\cite{DBLP:journals/csur/ZhuHM97, DBLP:journals/software/Runeson06, DBLP:journals/jcscoll/Olan03, DBLP:journals/pacmse/ChenZCWC25}.
However, during software evolution, tests can become outdated and incompatible with changes introduced to the system under test (SUT)~\cite{DBLP:conf/csmr/HurdugaciZ12, DBLP:conf/scam/MarsavinaRZ14, DBLP:conf/icst/ZaidmanRDD08, DBLP:conf/wcre/Wang000W21}, causing compilation or runtime failures.
Prior studies reported that broken test cases contribute to about one-fifth of failures in software systems~\cite{DBLP:conf/sigsoft/LabuschagneIH17, DBLP:conf/icsm/VahabzadehF015}.
False alarms raised by these broken tests \ie{failures caused by outdated test code rather than genuine SUT defects} can hinder the detection of real program faults~\cite{DBLP:conf/msr/LubsenZP09, DBLP:journals/corr/abs-2409-06416}.
Therefore, timely repair of unit tests is essential for software sustainability~\cite{DBLP:conf/csmr/HurdugaciZ12, DBLP:journals/corr/abs-2409-06416}.
Nevertheless, manually updating unit tests in response to SUT changes is time-consuming and error-prone~\cite{DBLP:journals/advse/KasurinenTS10, DBLP:conf/sigsoft/WidderHKV19, DBLP:conf/csmr/HurdugaciZ12, DBLP:series/springer/MoonenDZB08}, making automated test repair a practically important problem.

% 2. Existing work
% existing technical work
Automated \textbf{\textit{source--test co-evolution}} \ie{updating test code to match source code changes of the SUT} attracts much interest from researchers. 
Early source--test co-evolution techniques defined rules~\cite{DBLP:conf/kbse/DanielJDM09, DBLP:conf/icsm/MirzaAghaeiPP10, DBLP:conf/issta/DanielGM10, DBLP:conf/apsec/XuHWY14, DBLP:conf/sigsoft/MirzaAghaei11, DBLP:conf/icst/MirzaAghaeiPP12, DBLP:journals/stvr/MirzaAghaeiPP14, DBLP:conf/icst/LidO19} that apply predefined transformation templates for typical types of code changes. 
To overcome the rigidity and costly build of hand-crafted rules, later researchers built neural models for test repair \cite{DBLP:conf/kbse/HuLXLXY23,DBLP:journals/tse/YaraghiHKB25}, achieving broader applicability. 
Recently, large language models (LLMs) emerged as a more effective solution for test repair, showcasing promising performance due to their extensive code knowledge. 
They leverage context-aware test repair to tackle focal methods' signature changes \cite{DBLP:conf/issre/LiuYX0024}, utilize runtime feedback \cite{DBLP:journals/pacmse/ChiWHYCSS25, DBLP:conf/kbse/XuLYLXM25}, and address both test breakage and test enhancement \cite{DBLP:journals/pacmpl/RahmanKCGWMDR25, DBLP:journals/corr/abs-2509-24419}.
More recently, MuMuTestUp~\cite{DBLP:journals/pacmse/TianLPZCSS26} employs multiple LLM agents for mutation-guided repair of obsolete tests caused by the changes to a single focal method.\looseness=-1

% 3. Research gap
% technical gap
However, existing LLM-based test repair techniques primarily focus on scenarios involving changes to a single focal method, which is over-simplified compared to real-world co-evolution scenarios. 
By analyzing over 2,500 source--test co-evolution cases drawn from commits submitted within the past two years in highly-starred open-source Java projects, we characterize real-world co-evolution practices and observe that 70.9\% of broken tests invoke multiple changed methods and 77.8\% are accompanied by changes to invoked classes. 
These suggest that real-world source--test co-evolution is substantially more complex than the settings addressed by current techniques. 
Moreover, each technique proposes and evaluates one specific combination of repair context and pipeline design, leaving it unclear which design choices actually matter in such scenarios.

% existing empirical work & empirical gap
Existing empirical studies only partially answer this question.
Prior studies have characterized whether, when, and why production and test code co-evolve~\cite{DBLP:conf/icst/ZaidmanRDD08, DBLP:conf/sigsoft/PintoSO12, DBLP:journals/tosem/SunYLXLL23, DBLP:journals/tse/GuoCXCLZ25}, but not how broken tests should be repaired.
Very recent studies have begun to evaluate LLMs for test evolution, yet they either examine prompt design under single-focal-method changes~\cite{10.1145/3832282} or focus on the overall performance of coding agents~\cite{DBLP:journals/corr/abs-2605-06125, DBLP:journals/corr/abs-2607-02469}.

Therefore, the practical problem of how to enable LLMs in test repair in such complex, realistic scenarios remains underexplored. 
In particular, several fundamental design choices remain unsettled: 
what \textit{repair context} to supply; 
how the \textit{repair pipeline} should be orchestrated; 
whether the explicit reasoning steps taken by the LLM with reasoning mode enabled, which prove helpful in many coding tasks~\cite{yang2024chain, li2025structured, gao2025seer, yang2025think}, contribute to its superior performance over the LLM with reasoning mode disabled, and which steps are most effective;
how LLMs' repair performance varies across scenarios of \textit{increasing complexity}; and 
whether the derived insights can \textit{generalize to autonomous coding agents} and even further enhance their performance.

% 4. Our study and findings
In this paper, we aim to find the pivots of LLM-based test repair: the key design choices and model behaviors that determine whether a broken test can be successfully repaired in realistic source--test co-evolution scenarios. 
To answer the above questions, we conduct a systematic empirical study on a new benchmark comprising 500 real-world source--test co-evolution cases from 15 well-maintained Java projects on GitHub~\cite{GitHub}, covering commits between May 2024 and April 2026, which provides a more up-to-date and realistic evaluation with larger scale and more complex co-evolution scenarios than existing benchmarks. 
We investigate five research questions (RQs) regarding the key design dimensions of LLM setups for test repair:  
(RQ1) we assess how three types of \textit{repair context} \ie{Symptoms, Cause, and Support} affect test repair performance; 
(RQ2) we compare the effectiveness of popular \textit{pipeline orchestration} strategies \ie{in-context learning (ICL), multi-attempt repair, and iterative refinement} for test repair; 
(RQ3) we quantify the contribution of explicit reasoning by evaluating the same model with reasoning mode enabled and disabled, and establish a \textit{taxonomy} of reasoning actions, based on which we further distill the key reasoning steps to help LLMs repair test cost-effectively; 
(RQ4) we analyze how repair performance scales with repair scenario \textit{complexity} by partitioning benchmark instances based on the scope of code under test (CUT) changes and the adaptations required in the tests' golden patch; 
and (RQ5) we examine whether the preceding findings generalize to \textit{agent-based} test repair by measuring the improvement yield over a vanilla coding agent baseline. 

Our experimental results yield several insights. 
Specifically, we found that combining multiple context types is an effective strategy for LLM-based test repair. 
We also found that both multi-attempt repair and iterative refinement pipelines enhance LLMs' performance, while feedback from prior attempts can more efficiently guide the repair process, particularly during the initial self-check round. 
In addition, we derive a taxonomy of 14 common reasoning action patterns organized across four stages, revealing the reasoning behaviors that underlie the superior repair performance of the LLM with reasoning mode enabled over disabled. 
Based on this taxonomy, we successfully guide LLMs to repair tests effectively at a lower cost. 
Finally, our empirically validated findings translate into enhanced agent-based repair, achieving a 66.7\% relative improvement on challenging cases. 

% 7. Contributions
To summarize, this paper makes the following contributions:
\begin{itemize}[leftmargin=10pt]
    \item \textbf{Empirical Study}:
    We present a systematic empirical study on LLM-based test repair for source--test co-evolution, spanning repair context, pipeline orchestration, reasoning model behavior, scenario complexity, and generalization to the agent, organized by five research questions. 
    \item \textbf{Benchmark}:
    We construct a new benchmark for source--test co-evolution, comprising 500 dynamically validated, recent instances from 15 highly-starred open-source projects, addressing key limitations of existing benchmarks. 
    \item \textbf{Practical Guidelines}:
    We derive actionable guidelines for designing effective LLM-based test repair approaches, grounded in systematic empirical evidence across diverse co-evolution scenarios. We demonstrate that the findings can be translated into tangible improvements for agent-based test repair. 
\end{itemize}

\section{Background and Motivation}\label{sec:2}

\subsection{Test Repair for Source--Test Co-evolution}
During software evolution, source code changes can cause test cases to break, leading to compilation or runtime failures.
Test repair aims to automatically update broken test cases to restore their correctness and maintain the integrity of the test suite.
In this paper, we study test repair in the context of source–test co-evolution, where changes to the source code of a system under test (SUT) may necessitate corresponding updates to the test cases that verify its functionality.

Formally, given two versions of a SUT $V$ and $V'$, a test case $T$ that originally passed on $V$ may fail on $V'$ due to source code changes.
The goal of test repair is to produce a test case $T'$ that passes on $V'$ while preserving the original test intent:
\begin{equation}
    \begin{aligned}
        \text{TestRepair}(T, V, V') \rightarrow{T'}
    \end{aligned}
\end{equation}

\subsection{Motivating Example}
\label{subsec:motivatingexample}
\begin{figure*}[t!]
\centering
\includegraphics[width=\textwidth]{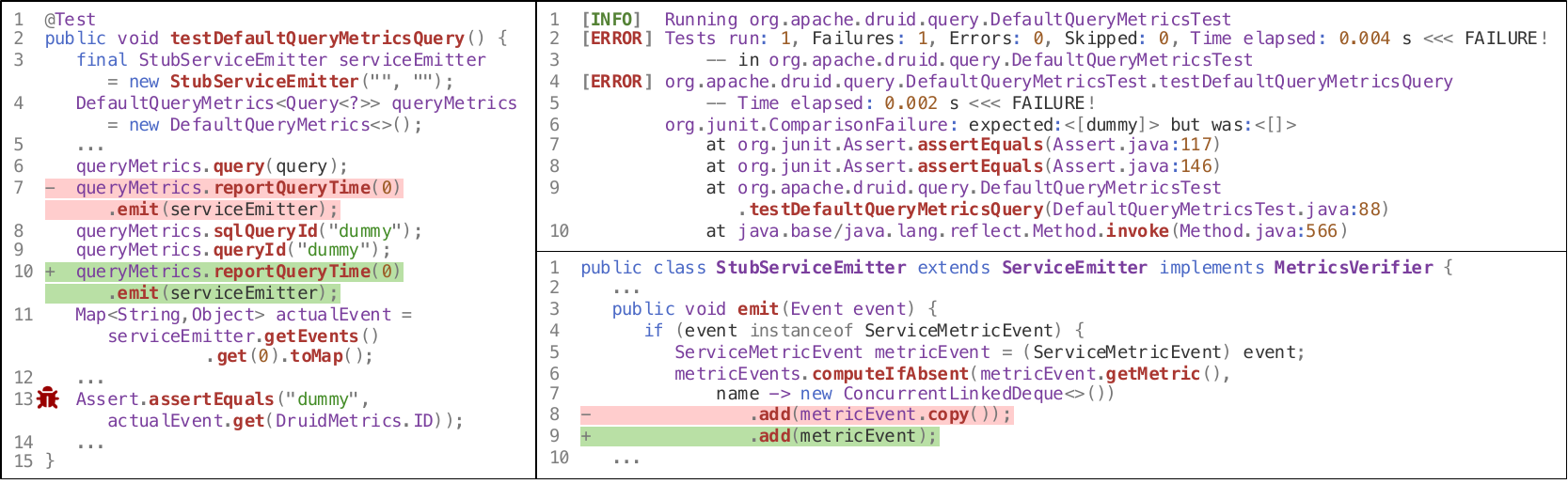}
\Description{A motivating example showing a source code change that breaks an existing test, the resulting test failure log, and the corresponding test repair that adapts the test to the evolved source code.}
\caption{Motivating example of test repair in source--test co-evolution. 
(Left: broken test with the patch submitted in commit \texttt{98c63de2cb}. Top-Right: failure log of the broken test. Bottom-right: relevant changes in the source code under test.)}
\label{motivating_example}
\vspace{-3mm}
\end{figure*}

We motivate the design of our research questions with a real-world source--test co-evolution example from \textit{apache/druid}, as shown in Figure~\ref{motivating_example}. 
An originally passing test, \code{testDefaultQueryMetricsQuery()}, fails with an \code{AssertionError} after commit \code{98c63de2cb} \cite{druidcommit98c63de2cb}, thus requiring repair. 
The ground truth test repair patch submitted by the developer in the same commit is shown in the left part of Figure~\ref{motivating_example}. 
However, generating this patch is non-trivial: it requires selecting the appropriate source code context to understand the failure, potentially refining the patch iteratively, and reasoning step by step to understand the root cause, before determining which statements in the test should be updated. 

Specifically, synthesizing the patch requires sufficient proper context to facilitate failure diagnosis. 
Commit \code{98c63de2cb} modified 18 methods in total, in which two methods are directly invoked by the broken test; however, neither of such tightly-related methods is responsible for the test breakage. 
The failure instead originates in class \code{StubServiceEmitter} (invoked transitively at line 3 of the test), which was changed to store the original event reference rather than a copy (\code{metricEvent} instead of \code{metricEvent.copy()}). 
To understand this failure, the failure log of the broken test is a helpful indicator. 
As shown in the right-top subgraph in Figure~\ref{motivating_example}, the log describes the failure \textit{Symptom}, locating where the test breaks \ie{line 88 in \code{DefaultQueryMetricsTest.java:88}} and what went wrong \ie{an \code{AssertionError} reporting a missing dimension value}.
In addition, the changes in the source code (under test) in the left subgraph explain the \textit{Cause} of the test breakage regarding what induced the test to fail only if captured at the proper granularity and scope. 
Note that the root cause of the failure is not in the directly-invoked methods, but rather in a transitive call to \code{StubServiceEmitter}, which is not immediately visible from the test case itself.
In addition, an unchanged method \code{ServiceEmitter.emit()} is also critical in this example to explain how things connect as \textit{Support} information. Serving as a dispatch mechanism, it connects the test case to the \code{StubServiceEmitter} class, which is where the actual failure occurs.
Together, these dimensions define a rich design space for repair context, motivating a systematic investigation of how \textit{Symptom}, \textit{Cause} (across hunk-, method-, and file-level granularities), and \textit{Support} context each affect repair performance and how they interact (\textbf{RQ1}). 

Beyond providing the proper context, the design of test repair pipeline may also facilitate the generation of correct test patches. 
For example, both iterative refinement and in-context learning (ICL) have been shown to enhance LLM performance in code generation tasks \cite{ICL1,ICL2,ICL0}. 
This motivates an investigation of how another attempt with the same context or feedback from the prior attempt, or additional test code examples can help LLMs produce a correct patch \textbf{(RQ2)}. 
In addition, the example also exposes demands of multiple reasoning steps \eg{integrating evidence across multiple changed files, understanding indirect runtime dispatch behavior, and inferring which statements in the test need to be updated}. 
This raises our interest to investigate whether the LLM with reasoning mode enabled that produces explicit chain-of-thought reasoning traces before generating code is better equipped for such scenarios than the LLM with reasoning mode disabled; and whether reasoning traces produced by the LLM with reasoning mode enabled can be leveraged to guide the LLM with reasoning mode disabled towards comparable repair performance at lower cost (\textbf{RQ3}). 

The example also motivates the investigation of repair performance across co-evolution cases with various complexity. 
For example, although this example involves only seven changed files, commits in real-world software evolution can touch hundreds of files. 
In our benchmark mined from recent two years' real-world software evolution, 12.6\% of the cases involve more than 100 changed Java files. 
The growing complexity of co-evolution scenarios motivates us to examine how repair performance varies across different contexts and still continue to hold in complex scenarios (\textbf{RQ4}). 
Finally, the expanding adoption of coding agents in software development steers us to ask whether these powerful agents can handle the complex cases where LLMs struggle, and whether the context strategies validated in our study can further improve their performance (\textbf{RQ5}).

\section{Study Design}\label{sec:3}
\subsection{Research Questions}
To systematically evaluate LLM-driven test repair performance for source--test co-evolution, we conduct experiments to address five research questions (RQs). 
These RQs investigate the impact of repair context types and pipeline orchestrations on repair performance, analyze the reasoning behaviors of the LLM with reasoning mode enabled, and explore how empirical findings can inform the design of more effective prompting strategies and repair pipelines following the motivating example in Section~\ref{subsec:motivatingexample}.

\begin{itemize}[leftmargin=10pt]
    \item \textbf{RQ1. How do different types of repair contexts affect test repair performance?} 
    We study three types of repair context as discussed in Section~\ref{subsec:motivatingexample}: (1) \textit{Symptom}, reflected by the test failure log \ie{the execution log of the broken test}; (2) \textit{Cause}, i.e., the changes in the code under test (CUT) at different levels of granularity \ie{hunk-level, method-level, and file-level}; and (3) \textit{Support}, which supplies unchanged but relevant methods and classes identified by LLMs. 

    \item \textbf{RQ2. How do different pipeline orchestrations affect test repair performance?} 
    We evaluate three popular pipeline orchestration strategies for LLMs, including in-context learning (ICL) \ie{prompting with $n$-shot example, $n{=}0, 1, 3$, retrieved based on similarity}, multi-attempt repair \ie{pass@$k$, $k{=}1, 2, 3$}, and iterative refinement \ie{$r$-iteration, $r{=}1, 2, 3$} to explore their impact on test repair performance. 

    \item \textbf{RQ3. What reasoning behaviors contribute to the superior performance of the LLM with reasoning mode enabled?} 
    We first examine the capability of the LLM with reasoning mode enabled for source--test co-evolution by comparing the performance of \textit{DeepSeek-V3.2}~\cite{deepseek-v3} with reasoning mode enabled versus disabled to understand the extent to which reasoning capabilities contribute to test repair. 
    Then, we analyze the reasoning traces produced by the LLM with reasoning mode enabled to identify recurring actions during test repair and distill these observations into a four-stage reasoning-action taxonomy. 
    We also investigate how the reasoning action taxonomy can inform the design of more cost-effective prompting strategies by regulating the repair pipeline to reduce unnecessary reasoning actions.

    \item \textbf{RQ4. How does the test repair performance scale with the repair complexity?} 
    We categorize complexity along two dimensions, namely \textit{the CUT change scale} \ie{the number of changed lines and methods in the CUT} and \textit{the test patch adaptation degree} \ie{the number of changed lines and hunks in the golden patch of the broken test}. 
    We then analyze the test repair performance for each category to understand how different strategies scale with increasing repair complexity. 

    \item \textbf{RQ5. Can the context configuration strategies studied in the previous RQs benefit agent-based test repair?} 
    We apply a SOTA coding agent, {Codex~\cite{codex}}, to the challenging repair tasks where simply prompting LLMs with test repair instructions \ie{the setup of RQs1--4} fail and systematically equip the agent with the context found helpful in our experiments of previous RQs. 
    Codex is equipped with a well-crafted harness to guide LLMs to iteratively and autonomously handle coding tasks. The comparison between its performance with and without our context demonstrates the usefulness of our findings in the more powerful agentic setting.
    
\end{itemize}

\subsection{Benchmark Construction}
\begin{table}[t]
    \small
    \centering
    \caption{Details of Projects and Samples in the Benchmark.}
    \setlength{\abovecaptionskip}{0pt}
    \label{tab:BenchmarkDetails}
    \begin{adjustbox}{max width=\linewidth}
    \begin{threeparttable}
    \begin{tabular}{l@{}r@{\quad}r@{ }r@{ }r@{ }r@{ }r}

    \toprule

    \makecell[c]{\textbf{Project}\\\textbf{Name}} & \makecell[c]{\textbf{Star}\\\textbf{Number}} & \makecell[c]{\textbf{Last}\\\textbf{Updated}} & \makecell[c]{\textbf{Commit}\\\textbf{Number}} & \makecell[c]{\textbf{\#Changed}\\\textbf{Tests}} & \makecell[c]{\textbf{\#Repaired}\\\textbf{Tests}} & \makecell[c]{\textbf{Sample}\\\textbf{Number}}\\

    \midrule

    \textit{shardingsphere}~\cite{shardingsphere} & 20,714 & 2026-05-10 & 6,339 & 855 & 720 & 134 \\

    \textit{druid}~\cite{druid}                  & 13,987 & 2026-05-10 & 1,900 & 902 & 574 & 107 \\

    \textit{iotdb}~\cite{iotdb}                  &  6,321 & 2026-05-11 & 3,091 & 1,206 & 497 & 95 \\

    \textit{spring-ai}~\cite{spring-ai} &  8,674 & 2026-05-10 & 2,881 & 998 & 198 & 37 \\

    \textit{pinpoint}~\cite{pinpoint}      & 13,820 & 2026-05-10 & 1,984 & 238 & 155 & 29 \\

    \textit{hudi}~\cite{hudi}                    &  6,158 & 2026-05-10 & 1,992 & 297 & 128 & 24 \\

    \textit{camel}~\cite{camel}                  &  6,199 & 2026-05-10 & 10,186 & 560 & 113 & 17 \\

    \textit{hertzbeat}~\cite{hertzbeat}           &  7,220 & 2026-05-11 & 1,340 & 124 &  72 & 13 \\

    \textit{spring-ai-alibaba}~\cite{spring-ai-alibaba} & 9,539 & 2026-05-11 & 3,137 & 136 & 45 & 12 \\

    \textit{quarkus}~\cite{quarkus}           & 15,657 & 2026-05-11 & 13,713 & 240 &  37 & 10 \\

    \textit{netty}~\cite{netty}                   & 34,944 & 2026-05-10 & 1,483 & 104 &  60 &  9 \\

    \textit{dubbo}~\cite{dubbo}                  & 41,526 & 2026-05-10 & 1,126 &  57 &  27 &  9 \\

    \textit{hazelcast}~\cite{hazelcast}       &  6,568 & 2026-05-10 & 2,285 & 115 &  32 &  2 \\

    \textit{DependencyCheck}~\cite{dependency-check} & 7,539 & 2026-05-10 & 1,117 & 39 & 16 & 1 \\

    \textit{jenkins}~\cite{jenkins}           & 25,257 & 2026-05-10 & 3,340 &  38 &   4 &  1 \\

    \midrule
    Total & & & & 5,909 & 2,678 & 500 \\

    \bottomrule
    \end{tabular}
    \end{threeparttable}
    \end{adjustbox}
    \vspace{-5mm}
\end{table}

Existing source--test co-evolution benchmarks suffer from several notable limitations. 
(1) \textit{Simplified source--test co-evolution scenarios.} 
Many benchmarks restrict code changes to overly simplified and aggressively filter out data. 
For instance, the SOTA benchmark Update4J~\cite{DBLP:journals/corr/abs-2509-24419} only includes broken tests caused by a single focal method change. 
Also, TaRBench~\cite{DBLP:journals/tse/YaraghiHKB25} only considers test repair patches with a single-hunk change. 
However, real-world test co-evolution often involves more complex change patterns. 
In our benchmark, 38.4\% of cases have test patches involving more than one changed hunk, which will be further discussed in RQ4 in Section~\ref{sec:4}. 
(2) \textit{Limited benchmark size.} 
The scale of existing benchmarks is relatively small \eg{Update4J~\cite{DBLP:journals/corr/abs-2509-24419} contains only 165 broken tests}, which limits the statistical power and generalizability of experimental results. 
(3) \textit{Test execution reproducibility.} 
Some benchmarks \eg{CEPROT~\cite{DBLP:conf/kbse/HuLXLXY23}} do not dynamically validate collected test cases, leaving downstream studies only able to reproduce and dynamically validate 0.44\% of the tests, raising concerns about benchmark reliability~\cite{DBLP:journals/pacmse/ChiWHYCSS25}. 
(4) \textit{Outdated data leading to data contamination.} 
Most commits in TaRBench~\cite{DBLP:journals/tse/YaraghiHKB25} were submitted before 2020. 
Even in the most recent benchmark, Update4J~\cite{DBLP:journals/corr/abs-2509-24419}, a substantial portion of the data predates 2020. 
Modern LLMs may have learned the correct test patches of these commits in their training data, inducing risks of data contamination. 

To address these limitations, we construct a new source--test co-evolution benchmark with a larger scale, more realistic code change scenarios, reliable test execution validation, and up-to-date data. 
The detailed data collection process and benchmark characteristics are described below. 

\textbf{Project Selection and Data Filtering.}
We collected commits from GitHub over a two-year window spanning from May~1, 2024 to April~30, 2026, ensuring that the data reflects recent development practices and partially postdates the training cutoff of modern LLMs, thereby mitigating data contamination. 
We selected Java projects managed by Maven~\cite{maven} and instrumented with JaCoCo~\cite{jacoco} to support test coverage recording. 
To focus on well-maintained repositories, we ranked candidates by GitHub stars and retained the top 500. 
We then removed the projects with fewer than 1,000 commits within the two-year window for insufficient evolution activities and kept 27 candidate projects after project filtering. 

For each candidate project, we identified genuine test repairs among test method changes following Yaraghi et al.~\cite{DBLP:journals/tse/YaraghiHKB25}, using a three-step dynamic validation procedure executed with Java~17 and Maven~3.9.9. 
Specifically, for every candidate test evolution instance in each project, we first executed the original test case $T$ against the original SUT version $V$ and verified that it passed to preserve the instances \textit{with test cases that initially pass}. 
Then, we executed $T$ against the modified SUT version $V'$ and checked whether it failed due to incompatibilities introduced by the SUT changes to keep the instances where \textit{SUT code changes break the original test Case}. 
Finally, we executed the modified test case $T'$ against $V'$ and verified that it passed, confirming a successful source--test co-evolution instance. 
After this three-step dynamic validation, we obtained 2,678 test cases from 15 projects.

From the pool of all dynamically validated co-evolution cases, we randomly sampled 500 instances to construct the final test repair benchmark. 
To the best of our knowledge, this benchmark is the first to capture large-scale, multi-method and multi-file source--test co-evolution cases with full dynamic validation from very recent software development practices. 
Detailed benchmark statistics are presented in Table~\ref{tab:BenchmarkDetails}.

\subsection{Experimental Setup}

\textbf{LLM Selection.}
We select one commercial LLM (\textit{GPT-5.4}~\cite{gpt5}) and two open-sourced LLMs (\textit{DeepSeek-V3.2}, \textit{abbr. DS-V3.2}~\cite{deepseek-v3} and \textit{Qwen3-Coder-480B-Instruct}, \textit{abbr. Qwen3-Coder}~\cite{qwen3-coder}) as experimental subjects since they show strong code generation capability and are built by different providers, demonstrating generalizable findings. 
Note that \textit{DeepSeek-V3.2} provides visible reasoning traces for analysis with configurable thinking capabilities when the reasoning mode is enabled. 
We require high thinking effort in RQ3 to investigate its most extensive reasoning traces. 
\textit{GPT-5.4} also supports configurable thinking efforts but does not provide visible reasoning traces, so we only use it with low thinking effort. 
We adopt the default configuration with temperature of 0.6, following the recommendation of the respective model vendors, to reflect realistic deployment conditions commonly used by end users~\cite{pan2024catcoder, halim2025studythinkingpatternslarge}.

\textbf{Evaluation Metrics.}
We adopt two widely-adopted metrics to quantify the effectiveness of test repair for source--test co-evolution following prior work on test repair~\cite{DBLP:journals/corr/abs-2509-24419, DBLP:conf/issre/LiuYX0024} and test generation~\cite{DBLP:conf/kbse/WangL0J24, DBLP:journals/pacmse/Yuan0DW00L24, DBLP:conf/sigsoft/ChenHZHDY24}:  
\begin{enumerate*}[label=(\arabic*)]
    \item{\textit{Compilation success rate (comp)}} measures the proportion of updated tests that compile without errors, capturing syntactic correctness and serving as a prerequisite for execution-based evaluation.
    \item{\textit{Test pass rate (exec)}} measures the proportion of updated tests that execute and pass against the updated source code, indicating whether the updated tests conform to the current system behavior. 
\end{enumerate*}

\textbf{Repair Quality Verification.}
A repaired test may trivially pass by weakening the test oracle \eg{removing assertions}.
To detect such degenerate repairs, we employ an LLM-as-a-Judge approach~\cite{zhou2025se, weyssow2026codeultrafeedback, pradel2025testora} with \textit{GPT-5.4} (thinking effort set to high), which compares each repair against the developer-committed test patch.
Note that the judge is designed to detect oracle weakening rather than to verify the full semantic correctness of a repair.
Across all experimental conditions, we generate a total of 92,054 test repairs. 
To validate the reliability of the judge, we randomly sample 383 repairs using the standard finite population sample size estimation formula~\cite{cochran1977sampling} with a 95\% confidence level and a margin of error of $\pm$5\%.
An author independently inspects each sampled repair to determine whether it weakens the test oracle, without access to the judge's verdict.
The human and LLM judgments agree on 98.17\% (375 out of 382) of the sampled repairs that only 1.87\% (7 out of 375) are oracle-weakening, indicating that the judge reliably identifies oracle-weakening repairs.
\section{Evaluation Results and Analysis}\label{sec:4}

\subsection{RQ1. Impact of Repair Context Types.}
\begin{table}[!t]
\setlength{\abovecaptionskip}{1pt}
\caption{Number and ratio of executable (\textit{exec}) and compilable (\textit{comp}) test repair generated by different LLMs across various context types over all 500 cases.}
\vspace{-1mm}
\setlength{\tabcolsep}{4pt}
\label{tab:RQ1}
\adjustbox{max width=\columnwidth, max totalheight=0.88\textheight}{%
\begin{tabular}{c|ll|llll|llll|llll}
\specialrule{\heavyrulewidth}{0pt}{0pt}
\multirow{2}{*}{\textbf{Context Type}} & \multicolumn{2}{c|}{\multirow{2}{*}{\textbf{Setups}}} & \multicolumn{4}{c|}{\textit{DeepSeek-V3.2}} & \multicolumn{4}{c|}{\textit{Qwen3-Coder}} & \multicolumn{4}{c}{\textit{GPT-5.4}} \\ \hhline{~~~*{12}{-}}
 & \multicolumn{2}{c|}{} & \multicolumn{2}{c}{\textit{exec}} & \multicolumn{2}{c|}{\textit{comp}} & \multicolumn{2}{c}{\textit{exec}} & \multicolumn{2}{c|}{\textit{comp}} & \multicolumn{2}{c}{\textit{exec}} & \multicolumn{2}{c}{\textit{comp}} \\ \specialrule{\lightrulewidth}{0pt}{0pt}
Symptom & \multicolumn{2}{l|}{\textit{Failure log}} & 88 & 0.176 & 149 & 0.298 & 115 & 0.230 & 181 & 0.362 & 157 & 0.314 & 226 & 0.452 \\ \hhline{*{15}{-}}
 & \multicolumn{2}{l|}{\textit{Hunk-level}} & 162 & 0.324 & 195 & 0.390 & 220 & 0.440 & 257 & 0.514 & 260 & 0.520 & 293 & 0.586 \\ \hhline{~*{14}{-}}
 & \multicolumn{2}{l|}{\textit{Method-level}} & 163 & 0.326 & 194 & 0.388 & 263 & 0.526 & 303 & 0.606 & 276 & 0.552 & 307 & 0.614 \\ \hhline{~*{14}{-}}
\multirow{-3}{*}{Cause} & \multicolumn{2}{l|}{\textit{File-level}} & \textbf{262} & \textbf{0.524} & \textbf{320} & \textbf{0.640} & \textbf{278} & \textbf{0.556} & \textbf{344} & \textbf{0.688} & \textbf{296} & \textbf{0.592} & \textbf{339} & \textbf{0.678} \\ \hhline{*{15}{-}}
 & \multicolumn{2}{l|}{\textit{Hunk-level + Failure log}} & 223 & 0.446 & 259 & 0.518 & 245 & 0.490 & 285 & 0.570 & 275 & 0.550 & 304 & 0.608 \\ \hhline{~*{14}{-}}
 & \multicolumn{2}{l|}{\textit{Method-level + Failure log}} & 218 & 0.436 & 256 & 0.512 & 273 & 0.546 & 308 & 0.616 & 283 & 0.566 & 316 & 0.632 \\ \hhline{~*{14}{-}}
\multirow{-3}{*}{\begin{tabular}[c]{@{}c@{}}Symptom +\\ Cause\end{tabular}} & \multicolumn{2}{l|}{\textit{File-level + Failure log}} & \textbf{304} & \textbf{0.608} & \textbf{354} & \textbf{0.708} & \textbf{293} & \textbf{0.586} & \textbf{351} & \textbf{0.702} & \textbf{321} & \textbf{0.642} & \textbf{364} & \textbf{0.728} \\ \hhline{*{15}{-}}
 & \multicolumn{1}{l|}{} & \textit{m=1} & 259 & 0.518 & 294 & 0.588 & 280 & 0.560 & 318 & 0.636 & 302 & 0.604 & 338 & 0.676 \\ \hhline{~~*{13}{-}}
 & \multicolumn{1}{l|}{} & \textit{m=3} & 273 & 0.546 & \textbf{315} & \textbf{0.630} & \textbf{288} & \textbf{0.576} & 323 & 0.646 & \textbf{324} & \textbf{0.648} & \textbf{356} & \textbf{0.712} \\ \hhline{~~*{13}{-}}
 & \multicolumn{1}{l|}{\multirow{-3}{*}{\textit{\begin{tabular}[c]{@{}l@{}}Hunk-level +\\ Failure log\end{tabular}}}} & \textit{m=5} & \textbf{274} & \textbf{0.548} & \textbf{315} & \textbf{0.630} & 283 & 0.566 & \textbf{324} & \textbf{0.648} & 320 & 0.640 & 356 & 0.712 \\ \hhline{~*{14}{-}}
 & \multicolumn{1}{l|}{} & \textit{m=1} & 257 & 0.514 & 298 & 0.596 & 283 & 0.566 & 324 & 0.648 & 308 & 0.616 & 351 & 0.702 \\ \hhline{~~*{13}{-}}
 & \multicolumn{1}{l|}{} & \textit{m=3} & 258 & 0.516 & 290 & 0.580 & \textbf{305} & \textbf{0.610} & \textbf{345} & \textbf{0.690} & \textbf{317} & \textbf{0.634} & 354 & 0.708 \\ \hhline{~~*{13}{-}}
 & \multicolumn{1}{l|}{\multirow{-3}{*}{\textit{\begin{tabular}[c]{@{}l@{}}Method-level +\\ Failure log\end{tabular}}}} & \textit{m=5} & \textbf{266} & \textbf{0.532} & \textbf{301} & \textbf{0.602} & 294 & 0.588 & 336 & 0.672 & 315 & 0.630 & \textbf{359} & \textbf{0.718} \\ \hhline{~*{14}{-}}
 & \multicolumn{1}{l|}{} & \textit{m=1} & 298 & 0.596 & 346 & 0.692 & 301 & 0.602 & 361 & 0.722 & 341 & 0.682 & 386 & 0.772 \\ \hhline{~~*{13}{-}}
 & \multicolumn{1}{l|}{} & \textit{m=3} & 305 & 0.610 & 352 & 0.704 & \textbf{313} & \textbf{0.626} & \textbf{374} & \textbf{0.748} & 339 & 0.678 & 385 & 0.770 \\ \hhline{~~*{13}{-}}
\multirow{-9}{*}{\begin{tabular}[c]{@{}c@{}}Symptom +\\ Cause +\\ Top-$m$ Support\end{tabular}} & \multicolumn{1}{l|}{\multirow{-3}{*}{\textit{\begin{tabular}[c]{@{}l@{}}File-level +\\ Failure log\end{tabular}}}} & \textit{m=5} & \textbf{311} & \textbf{0.622} & \textbf{357} & \textbf{0.714} & 301 & 0.602 & 359 & 0.718 & \textbf{345} & \textbf{0.690} & \textbf{391} & \textbf{0.782} \\ \specialrule{\heavyrulewidth}{0pt}{0pt}
\end{tabular}%
}
\vspace{-5mm}
\end{table}

In RQ1, we study how \textit{Symptom}, \textit{Cause}, and \textit{Support} contexts contribute to LLMs' test repair performance.
Specifically, as discussed in Section~\ref{subsec:motivatingexample}, we collect failure logs of broken tests as the {{\textit{Symptom}}} context. 
To represent {\textit{Cause}} context, we generate coverage reports using JaCoCo and retrieve hunk-, method-, and file-level differences based on the commit changes tracked by Git~\cite{Git}.
For {\textit{Support}} context, we follow recent test update work~\cite{DBLP:journals/corr/abs-2509-24419, DBLP:journals/pacmse/ChiWHYCSS25} to prompt a frontier LLM \ie{\textit{DeepSeek-V3.2}} to identify the top-$m$ ($m{=}1,3,5$) relevant unchanged methods and classes from the SUT based on the broken test, the failure log, and fully qualified names of methods and classes in the CUT changes. 
We then retrieve their code bodies as context via a language server. 
We also combine multiple context types to evaluate their complementary effects.

Table~\ref{tab:RQ1} presents the overall results of different models' test repair performance on our benchmark across various context configurations. 
For each of the three LLMs, we present its \textit{exec} and \textit{comp} of generated repairs, providing only Symptom (row 1), only Cause (rows 2-4), Symptom and Cause combined (rows 5-7), and combinations of three context types (rows 8-16).
We derive two key observations from the results as follows.

\textit{\textbf{Three context types complement each other and jointly improve the overall repair performance.}} 
As shown in Table~\ref{tab:RQ1}, incorporating all three context types (Symptom\,+\,Cause\,+\,Top-$m$ Support) consistently yields the best performance across all three models. 
The adoption of both Symptom and Cause (Symptom\,+\,Cause) also improves performance compared to using only Symptom or Cause, indicating that each context type provides complementary information for test repair.
Specifically, when only the test failure log is provided (Symptom), performance is low \eg{23.0\% \textit{exec} and 36.2\% \textit{comp} using \textit{Qwen3-Coder}}. 
However, combining the failure log with CUT changes (Symptom\,+\,Cause) significantly improves the performance, increasing \textit{Qwen3-Coder}'s \textit{exec} by 113\% with hunk-level CUT changes, 137\% with method-level CUT changes, and 155\% with file-level CUT changes. 
Although the performance of all three LLMs with only Cause context is already high, incorporating Symptom further improves \textit{exec} by 16.0\% for \textit{DeepSeek-V3.2}, 5.40\% for \textit{Qwen3-Coder}, and 8.45\% for \textit{GPT-5.4}. 
This indicates that the symptom in failure logs, while insufficient on its own, helps the model pinpoint the relevant portion of the production code change. 
We consider the reason is that each context type captures a distinct aspect of the repair signal: 
Symptom reveals \textit{what} and \textit{where} failed, Cause reveals \textit{why} it failed, and Support provides additional code context, making them naturally complementary. 

\textit{\textbf{More context is not always better. Reducing redundancy across code context and matching model capacity are the keys to successful repair.}} 
We observe that two factors \ie{code context's granularity diversity and model's capacity} play a critical role in determining the optimal context configuration. 
First, after incorporating Support context (Symptom\,+\,Cause\,+\,Top-$m$ Support), \textit{DeepSeek-V3.2} and \textit{GPT-5.4} perform worse with method-level CUT changes than with hunk-level, and the advantage of method-level over hunk-level for \textit{Qwen3-Coder} drops greatly \eg{from 11.4\% to 1.07\% in \textit{exec} with top-1 support}. 
These observations exhibit a general reversal of the granularity trend seen without Support context (Symptom\,+\,Cause), likely because the method- and class-level code bodies in Support context overlap with the structural information that already supplied by method-level CUT changes, inducing redundancy that dilutes the model's focus on the core change. 
In contrast, hunk-level CUT changes capture a finer-grained code change perspective structurally distinct from Support context, being complementary. 
This suggests that \textit{code context's granularity diversity} across context dimensions, rather than sheer volume, is significant to performance gains. 
Second, the optimal amount of Support context varies with model capacity.
\textit{Qwen3-Coder} achieves its best \textit{exec} at top-$3$ support across all CUT change granularity levels, with top-$5$ providing no further gain or even a slight regression (Symptom\,+\,Cause\,+\,Top-$m$ Support, $m$=3, 5).
In contrast, \textit{GPT-5.4} that has a larger context window continues to benefit from top-$5$ support (Symptom\,+\,Cause\,+\,Top-5 Support).
This suggests that models with smaller effective context utilization capacity may suffer from information overload when excessive support context is supplied, and that the optimal $m$ should be calibrated to the \textit{model's capacity}.

\begin{tcolorbox}[boxsep=2pt,left=2pt,right=2pt,top=2pt,bottom=2pt,width=\linewidth,colback=white!90!black,boxrule=0pt, colbacktitle=white!,toptitle=2pt,bottomtitle=1pt,opacitybacktitle=0]
\textbf{Answer to RQ1.}
The three context types are complementary, and their combination offers better performance. 
Yet inclusion of more context is not always an advantage: code context redundancy and the model's capacity to utilize context are critical to the optimal context configuration. 
\end{tcolorbox}

\subsection{RQ2. Impact of Pipeline Orchestrations}
\begin{figure}[t!]
\centering
\renewcommand{\thesubfigure}{%
  \ifcase\value{subfigure}%
  \or a1-1%
  \or a1-2%
  \or a1-3%
  \or a2-1%
  \or a2-2%
  \or a2-3%
  \or a3-1%
  \or a3-2%
  \or a3-3%
  \or a4-1%
  \or a4-2%
  \or a4-3%
  \fi}

\begin{minipage}[t]{0.32\linewidth}
    \centering
    % \vspace{-1mm}
    \includegraphics[width=\linewidth]{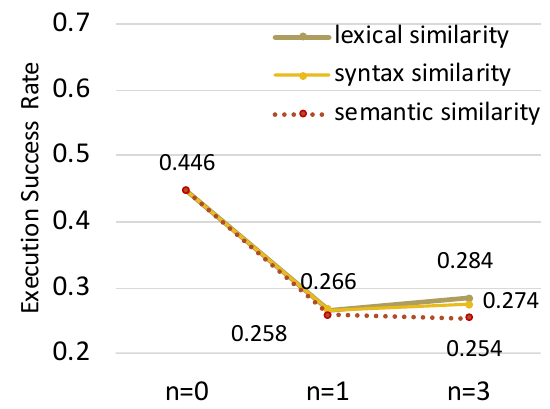}
    \Description{Line chart of execution success rate under in-context learning with varying numbers of shots, given hunk-level source change context and failure log.}
    \subcaption{ICL ($n$-shot) with hunk-level CUT changes and log (\textit{DS-V3.2}).}
    \label{fig:RQ2_a1}
\end{minipage}
\hfill
\begin{minipage}[t]{0.32\linewidth}
    \centering
    % \vspace{-1mm}
    \includegraphics[width=\linewidth]{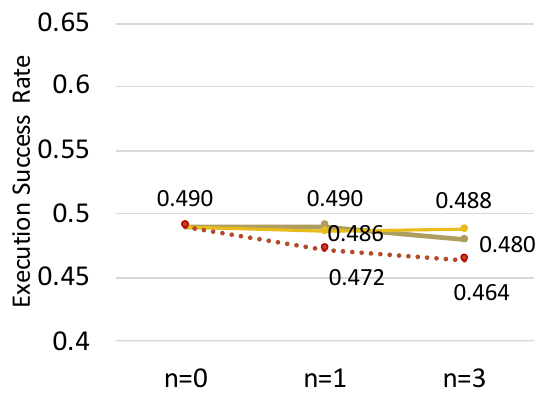}
    \Description{Line chart of execution success rate under in-context learning with varying numbers of shots, given method-level source change context and failure log.}
    \subcaption{ICL ($n$-shot) with hunk-level CUT changes and log (\textit{Qwen3-Coder}).}
    \label{fig:RQ2_a2}
\end{minipage}
\hfill
\begin{minipage}[t]{0.32\linewidth}
    \centering
    % \vspace{-1mm}
    \includegraphics[width=\linewidth]{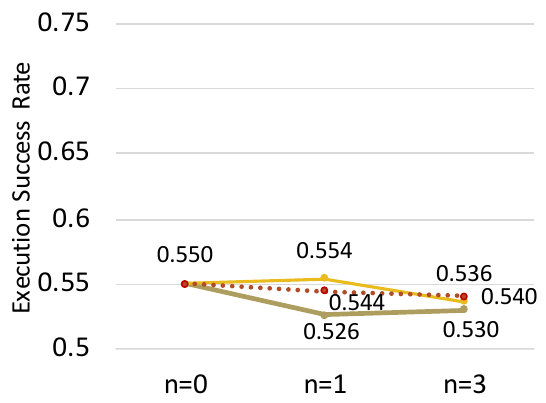}
    \Description{Line chart of execution success rate when the model makes multiple independent repair attempts, shown for increasing numbers of attempts.}
    \subcaption{ICL ($n$-shot) with hunk-level CUT changes and log (\textit{GPT-5.4}).}
    \label{fig:RQ2_b}
\end{minipage}

\begin{minipage}[t]{0.32\linewidth}
    \centering
    % \vspace{-1mm}
    \includegraphics[width=\linewidth]{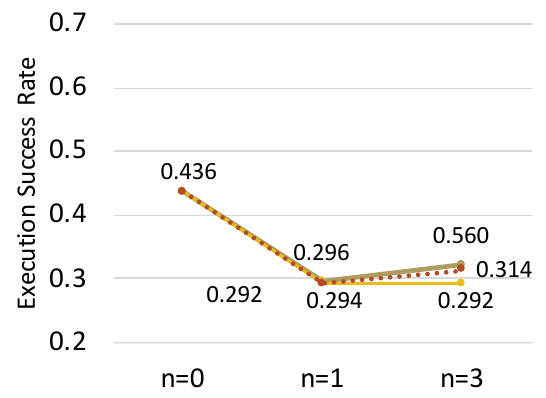}
    \Description{Line chart of execution success rate under in-context learning with varying numbers of shots, given file-level source change context and failure log.}
    \subcaption{ICL ($n$-shot) with method-level CUT changes and log (\textit{DS-V3.2}).}
    \label{fig:RQ2_a3}
\end{minipage}
\hfill
\begin{minipage}[t]{0.32\linewidth}
    \centering
    % \vspace{-1mm}
    \includegraphics[width=\linewidth]{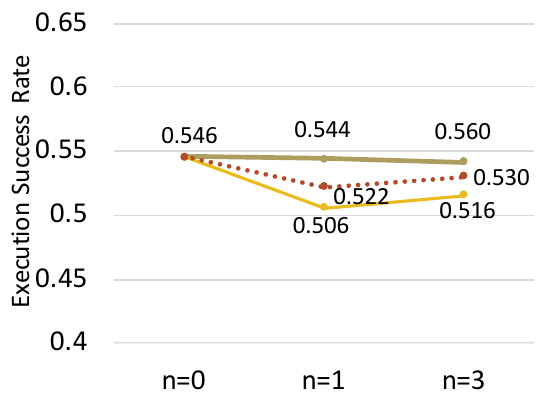}
    \Description{Line chart of execution success rate under in-context learning with varying numbers of shots, given file-level source change context, failure log, and top-5 supporting context.}
    \subcaption{ICL ($n$-shot) with method-level CUT changes and log (\textit{Qwen3-Coder}).}
    \label{fig:RQ2_a4}
\end{minipage}
\hfill
\begin{minipage}[t]{0.32\linewidth}
    \centering
    % \vspace{-1mm}
    \includegraphics[width=\linewidth]{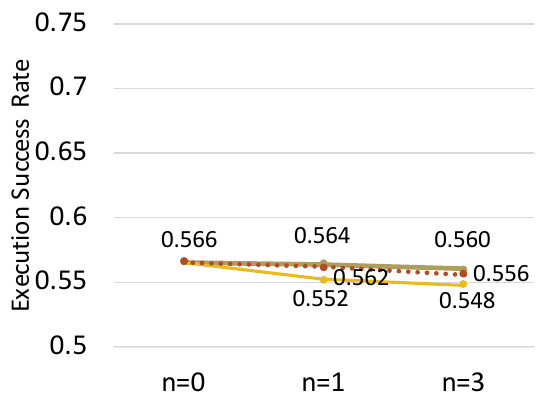}
    \Description{Line chart of execution success rate under iterative refinement, shown for increasing numbers of repair rounds.}
    \subcaption{ICL ($n$-shot) with method-level CUT changes and log (\textit{GPT-5.4}).}
    \label{fig:RQ2_c}
\end{minipage}
\hfill

\begin{minipage}[t]{0.32\linewidth}
    \centering
    % \vspace{-1mm}
    \includegraphics[width=\linewidth]{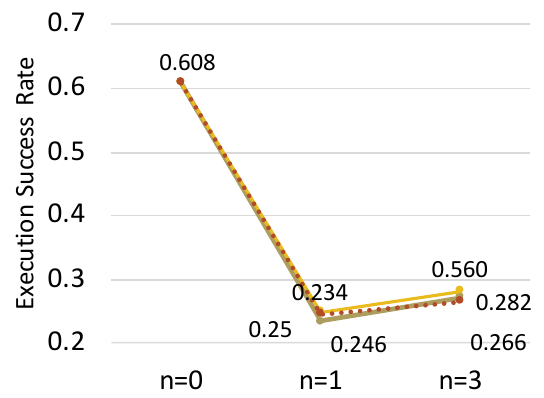}
    \Description{Line chart of execution success rate under iterative refinement, shown for increasing numbers of repair rounds.}
    \subcaption{ICL ($n$-shot) with file-level CUT changes and log (\textit{DS-V3.2}).}
    \label{fig:RQ2_c2}
\end{minipage}
\hfill
\begin{minipage}[t]{0.32\linewidth}
    \centering
    % \vspace{-1mm}
    \includegraphics[width=\linewidth]{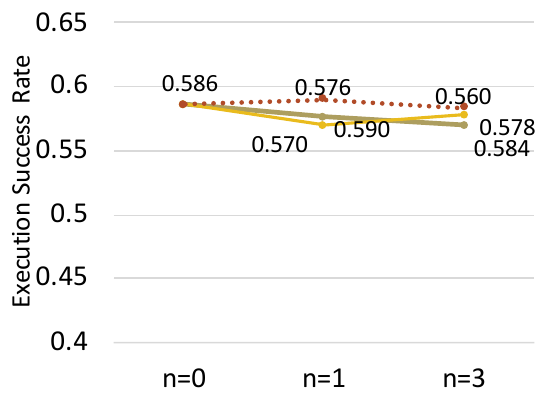}
    \Description{Line chart of execution success rate under iterative refinement, shown for increasing numbers of repair rounds.}
    \subcaption{ICL ($n$-shot) with file-level CUT changes and log (\textit{Qwen3-Coder}).}
    \label{fig:RQ2_c3}
\end{minipage}
\hfill
\begin{minipage}[t]{0.32\linewidth}
    \centering
    % \vspace{-1mm}
    \includegraphics[width=\linewidth]{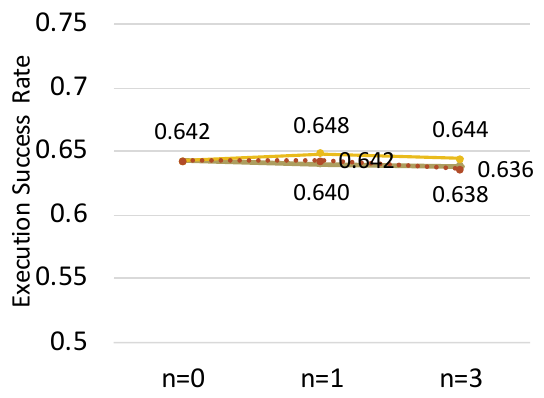}
    \Description{Line chart of execution success rate under iterative refinement, shown for increasing numbers of repair rounds.}
    \subcaption{ICL ($n$-shot) with file-level CUT changes and log (\textit{GPT-5.4}).}
    \label{fig:RQ2_c4}
\end{minipage}
\hfill

\begin{minipage}[t]{0.32\linewidth}
    \centering
    % \vspace{-1mm}
    \includegraphics[width=\linewidth]{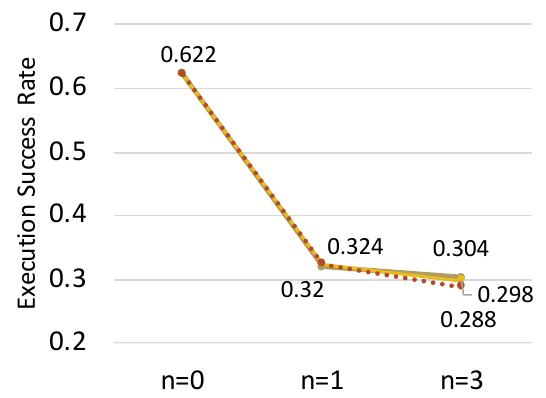}
    \Description{Line chart of execution success rate under iterative refinement, shown for increasing numbers of repair rounds.}
    \subcaption{ICL ($n$-shot) with file-level CUT changes, log, top-5 support (\textit{DS-V3.2}).}
    \label{fig:RQ2_c5}
\end{minipage}
\hfill
\begin{minipage}[t]{0.32\linewidth}
    \centering
    % \vspace{-1mm}
    \includegraphics[width=\linewidth]{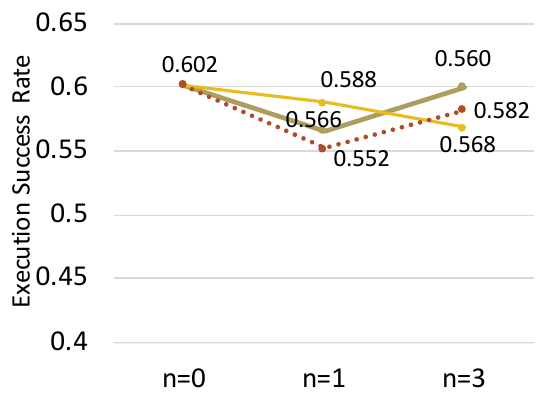}
    \Description{Line chart of execution success rate under iterative refinement, shown for increasing numbers of repair rounds.}
    \subcaption{ICL ($n$-shot) with file-level CUT changes, log, top-5 support (\textit{Qwen3-Coder}).}
    \label{fig:RQ2_c6}
\end{minipage}
\hfill
\begin{minipage}[t]{0.32\linewidth}
    \centering
    % \vspace{-1mm}
    \includegraphics[width=\linewidth]{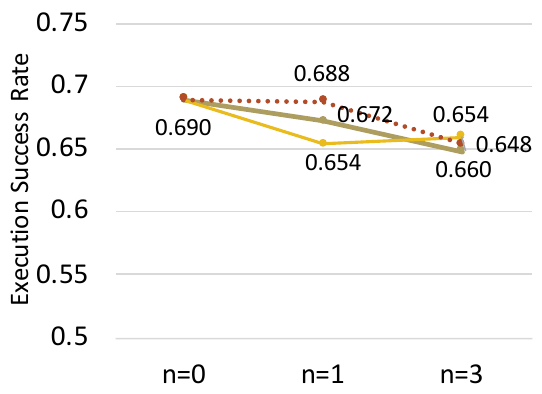}
    \Description{Line chart of execution success rate under iterative refinement, shown for increasing numbers of repair rounds.}
    \subcaption{ICL ($n$-shot) with file-level CUT changes, log, top-5 support (\textit{GPT-5.4}).}
    \label{fig:RQ2_c7}
\end{minipage}

\setlength{\abovecaptionskip}{3pt}
\vspace{1mm}
\caption{Execution success rate (\textit{exec}) of tests repaired by different LLMs with ICL under different context configurations.
Panels (a1-1)-(a1-3), (a2-1)-(a2-3), (a3-1)-(a3-3), and (a4-1)-(a4-3) correspond to Hunk-level\,+\,Failure log, Method-level\,+\,Failure log, File-level\,+\,Failure log, and File-level\,+\,Failure log\,+\,Top-5 Support respectively.}

\label{fig:RQ2_ICL}
\vspace{-3mm}
\end{figure}

\begin{figure}[t!]
% \vspace{-1mm}
\centering
\renewcommand{\thesubfigure}{%
  \ifcase\value{subfigure}%
  \or b1%
  \or b2%
  \or b3%
  \fi}

\begin{minipage}[t]{0.32\linewidth}
    \centering
    % \vspace{-1mm}
    \includegraphics[width=\linewidth]{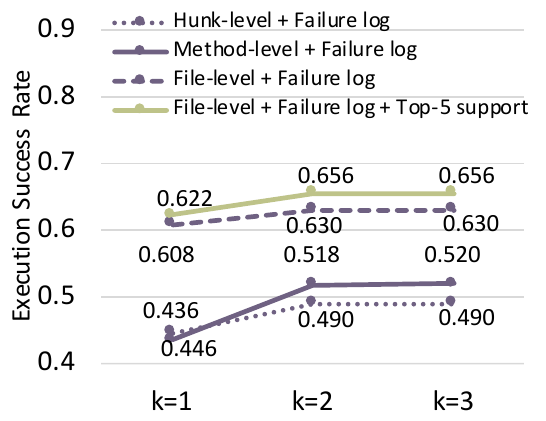}
    \Description{Line chart of execution success rate under iterative refinement, shown for increasing numbers of repair rounds.}
    \subcaption{Repair with $k$ attempts (\textit{DS-V3.2}).}
    \label{fig:RQ2_c8}
\end{minipage}
\hfill
\begin{minipage}[t]{0.32\linewidth}
    \centering
    % \vspace{-1mm}
    \includegraphics[width=\linewidth]{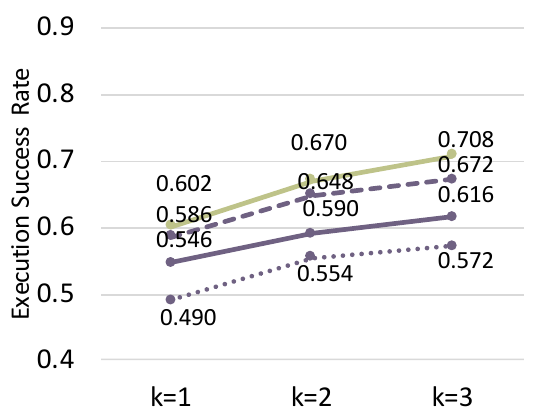}
    \Description{Line chart of execution success rate under iterative refinement, shown for increasing numbers of repair rounds.}
    \subcaption{Repair with $k$ attempts (\textit{Qwen3-Coder})}
    \label{fig:RQ2_c9}
\end{minipage}
\hfill
\begin{minipage}[t]{0.32\linewidth}
    \centering
    % \vspace{-1mm}
    \includegraphics[width=\linewidth]{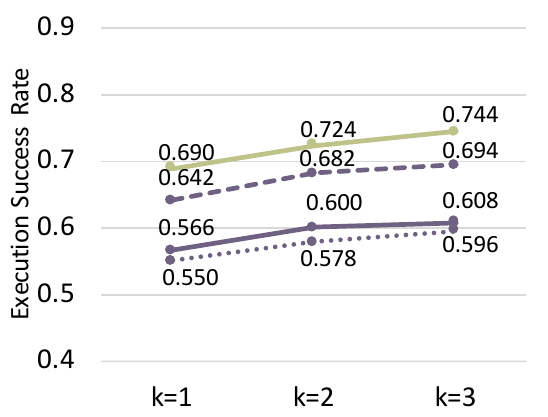}
    \Description{Line chart of execution success rate under iterative refinement, shown for increasing numbers of repair rounds.}
    \subcaption{Repair with $k$ attempts (\textit{GPT-5.4})}
    \label{fig:RQ2_c10}
\end{minipage}

\setlength{\abovecaptionskip}{3pt}
\caption{Execution success rate (\textit{exec}) of tests repaired by different LLMs with multiple attempts under different context configurations.
}
\label{fig:RQ2_multiAttempt}
\vspace{-1mm}
\end{figure}

\begin{figure}[t!]
\centering
\renewcommand{\thesubfigure}{%
  \ifcase\value{subfigure}%
  \or c1%
  \or c2%
  \or c3%
  \fi}

\begin{minipage}[t]{0.32\linewidth}
    \centering
    % \vspace{-1mm}
    \includegraphics[width=\linewidth]{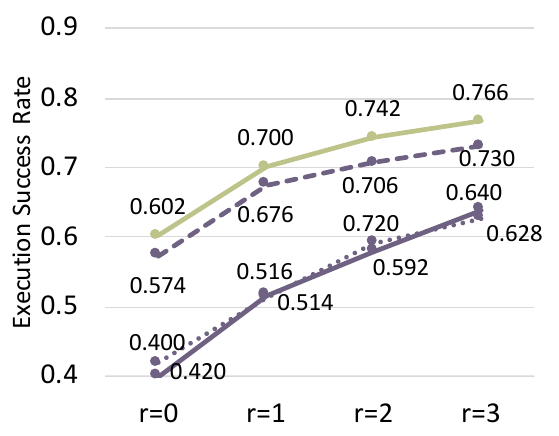}
    \Description{Line chart of execution success rate under iterative refinement, shown for increasing numbers of repair rounds.}
    \subcaption{Repair with $r$ iterative rounds (\textit{DS-V3.2}).}
    \label{fig:RQ2_c11}
\end{minipage}
\hfill
\begin{minipage}[t]{0.32\linewidth}
    \centering
    % \vspace{-1mm}
    \includegraphics[width=\linewidth]{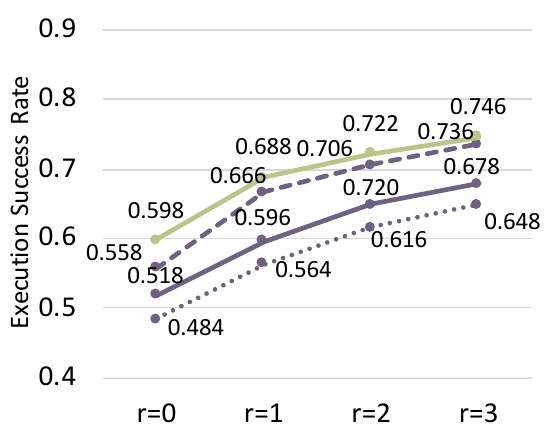}
    \Description{Line chart of execution success rate under iterative refinement, shown for increasing numbers of repair rounds.}
    \subcaption{Repair with $r$ iterative rounds (\textit{Qwen3-Coder}).}
    \label{fig:RQ2_c12}
\end{minipage}
\hfill
\begin{minipage}[t]{0.32\linewidth}
    \centering
    % \vspace{-1mm}
    \includegraphics[width=\linewidth]{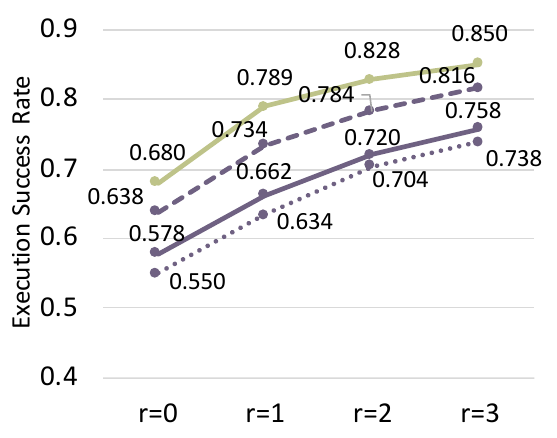}
    \Description{Line chart of execution success rate under iterative refinement, shown for increasing numbers of repair rounds.}
    \subcaption{Repair with $r$ iterative rounds (\textit{GPT-5.4}).}
    \label{fig:RQ2_c13}
\end{minipage}

\setlength{\abovecaptionskip}{3pt}
\caption{Execution success rate (\textit{exec}) of tests repaired by different LLMs with iterative rounds under different context configurations.
}
\label{fig:RQ2_itRepair}
\vspace{-2mm}
\end{figure}

In RQ2, we evaluate LLM performance with three pipeline designs, namely \textit{In-context learning (ICL)}, \textit{Multi-attempt repair} and \textit{Iterative refinement}. 
We evaluate the three LLMs \ie{\textit{DeepSeek-V3.2}, \textit{Qwen3-Coder}, \textit{GPT-5.4}} under the optimal context configuration \ie{Symptom\,+\,Cause at file-level\,+\,Top-5 Support} together with context configurations with three CUT change granularities \ie{Symptom\,+\,Cause at hunk-/method-/file-level}.

For ICL, we provide $n$-shot examples ($n{{=}0, 1, 3}$) of tests retrieved from version $V$ in the prompt to guide the repair, following the common practice for ICL-based coding tasks~\cite{DBLP:conf/icse/NashidSM23, ICL0, ICL1, ICL2}. 
We rank candidate tests by similarity to the broken test using three popular combinations of {source representation method} and {comparison method}:
(1) \textit{Lexical}, which represents tests as token bags via a PL-based tokenizer and ranks with BM25~\cite{DBLP:journals/ftir/RobertsonZ09}; 
(2) \textit{Syntax}, which encodes tests as abstract syntax trees (ASTs) and computes similarity via the Jaccard index~\cite{DBLP:conf/icse/NashidSM23}; 
and (3) \textit{Semantic}, which encodes tests using CodeBERT~\cite{DBLP:conf/emnlp/FengGTDFGS0LJZ20} and ranks by cosine similarity. 
For \textit{Multi-attempt repair}, we generate up to $k$ repair attempts ($k{{=}1, 2, 3}$) for each test, and consider the repair successful if any attempt executes successfully. 
For \textit{Iterative refinement}, we refine prior repair up to $r$ iterations ($r{{=}0, 1, 2, 3}$) by feeding back the execution result and failure log from each prior attempt, stopping upon a successful repair or when $r{{=}3}$. 

Figures~\ref{fig:RQ2_ICL}, \ref{fig:RQ2_multiAttempt}, and \ref{fig:RQ2_itRepair} illustrate the performance of the three LLMs with ICL, multi-attempt repair, and iterative refinement, respectively, across various context configurations.
As shown in Figure~\ref{fig:RQ2_ICL}, surprisingly, ICL with few-shot examples does not provide additional benefits over zero-shot repair across all three LLMs and all four context configurations, and sometimes even leads to performance degradation.
This suggests that test examples retrieved from version $V$ may not serve as effective repair demonstrations, despite sharing similar code features with the broken test. 
A likely explanation is that in source--test co-evolution, the broken test and its retrieved examples from version $V$ share the same outdated patterns: in source--test co-evolution practice, similar test patches are often applied in a batch \ie{in the same commit submission}. 
Although they may help code generation within the same code version, these examples tend to co-evolve together with the broken test and thus introduce noise rather than useful guidance.
Notably, the degradation is particularly severe for \textit{DeepSeek-V3.2}, whose \textit{exec} drops sharply once few-shot examples are added (e.g., from 0.608 to around 0.25 with File-level\,+\,Failure log), whereas \textit{Qwen3-Coder} and \textit{GPT-5.4} remain relatively stable.
This might because \textit{DeepSeek-V3.2} relies more heavily on in-context examples, imitating the outdated code patterns in the retrieved examples rather than adapting the broken test to the changed CUT. 

Meanwhile, multi-attempt repair (Figure~\ref{fig:RQ2_multiAttempt}) yields steady improvements across all the three LLMs and all context settings, as more independent attempts increase the chance of a passing repair. 
Iterative refinement achieves larger gains for all the three LLMs, most pronounced in the first iteration, as shown in Figure~\ref{fig:RQ2_itRepair}. 
This suggests that properly collecting and utilizing initial repair attempt's feedback for further improvement is more effective than simply generating a second independent attempt. 

\vspace{-1mm}
\begin{tcolorbox}[boxsep=2pt,left=2pt,right=2pt,top=2pt,bottom=2pt,width=\linewidth,colback=white!90!black,boxrule=0pt, colbacktitle=white!,toptitle=2pt,bottomtitle=1pt,opacitybacktitle=0]
\textbf{Answer to RQ2.} 
ICL with few shots retrieved from SUT $V$ does not improve repair performance and may even degrade it, whereas multi-attempt repair and iterative refinement both help. 
Iterative refinement brings substantial gains in the first round, suggesting that feedback from initial attempts is crucial to repair effectiveness. 
\end{tcolorbox}

\subsection{RQ3. Performance and Behavior Investigation for the LLM with Reasoning Mode Enabled}

In RQ3, we first compare the performance of \textit{DeepSeek-V3.2} with reasoning mode enabled (thinking effort set to \textit{high}) versus disabled across four context configurations (same as RQ2) to quantify the impact of reasoning capabilities on test repair. 
We focus on \textit{DeepSeek-V3.2} first since it is the only open-source LLM among the three that offers configurable reasoning ability and accessible reasoning traces. 
Next, we analyze the reasoning traces produced by \textit{DeepSeek-V3.2} with reasoning mode enabled during successful repairs to identify recurring behaviors and distill them into a structured taxonomy. 
Finally, we derive two optimization strategies based on the reasoning behavior taxonomy. 
In order to evaluate the generalizability, we apply the two optimization strategies to all three LLMs \ie{\textit{DeepSeek-V3.2}, \textit{Qwen3-Coder}, \textit{GPT-5.4}} in terms of both repair effectiveness and cost efficiency.   

\begin{figure}[t!]
\centering
% \vspace{-3mm}
\begin{minipage}[t]{0.495\linewidth}
    \centering
    \includegraphics[width=\linewidth]{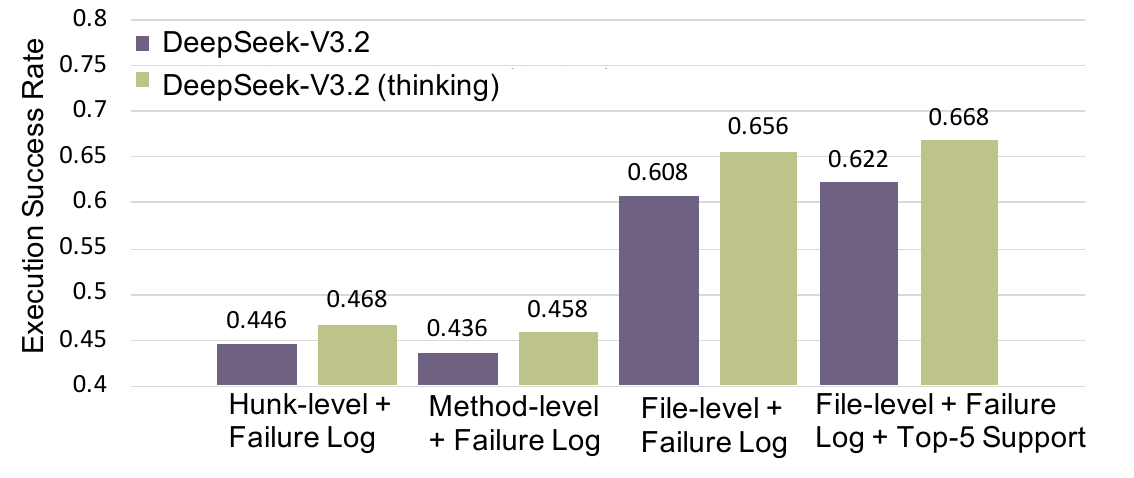}
    \Description{Bar chart comparing the execution success rate of large language models with thinking mode enabled versus disabled across different context configurations.}
    \subcaption{Execution success rate (\textit{exec}).}
    \label{fig:RQ3_exec}
\end{minipage}
\hfill
\begin{minipage}[t]{0.495\linewidth}
    \centering
    \includegraphics[width=\linewidth]{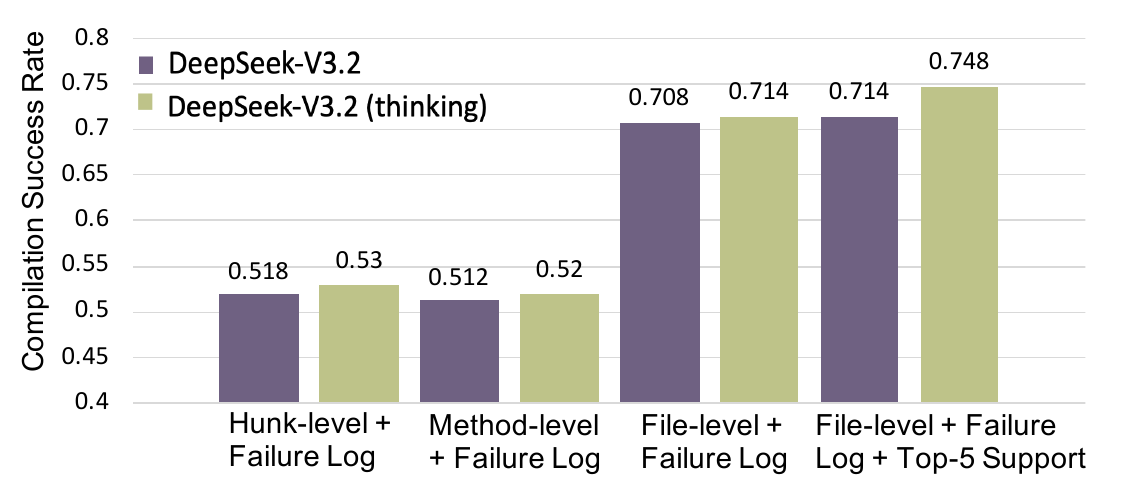}
    \Description{Bar chart comparing the compilation success rate of large language models with thinking mode enabled versus disabled across different context configurations.}
    \subcaption{Compilation success rate (\textit{comp}).}
    \label{fig:RQ3_comp}
\end{minipage}
% \vspace{-3mm}
\caption{Comparison of LLM with thinking mode enabled versus disabled on test repair performance.}
\label{fig:RQ3}
\vspace{-5mm}
\end{figure}

\textit{\textbf{The model with reasoning mode enabled consistently outperforms the model with reasoning mode disabled across all context configurations.}} 
Figure~\ref{fig:RQ3} presents the experimental results, with detailed statistics provided in Table~\ref{tab:RQ34} (rows of model \textit{DS-V3.2} and \textit{DS-V3.2 (thinking)} with baseline optimization setup). 
We observe that enabling reasoning mode yields consistent improvements in both \textit{exec} and \textit{comp} across all context configurations. 
It achieves the largest \textit{exec} gain of 7.89\% with \textit{File-level\,+\,Failure log} context, and the smallest \textit{exec} gain of 4.93\% with \textit{Hunk-level\,+\,Failure log} context. 

\textit{\textbf{Successful repairs by the LLM with reasoning mode enabled follow a four-stage reasoning process comprising 14 common reasoning action patterns.}}
To understand \textit{how} LLM with reasoning mode enabled succeed at test repair, we manually analyzed reasoning traces of successfully repaired cases using the open coding protocol~\cite{khandkar2009open}, following Halim et al.~\cite{halim2025studythinkingpatternslarge}. 
Two annotators with over eight years of programming experience independently labeled fine-grained reasoning behaviors in a random sample of traces, then iteratively grouped similar labels and resolved disagreements through discussion with an arbitrator until consensus was reached. 
Finally, we develop a reasoning action taxonomy comprising four \textit{stages} with 14 common reasoning \textit{action patterns}. 

\begin{table*}[t!]
\setlength{\abovecaptionskip}{0pt}
\caption{Statistics of reasoning action taxonomy. 
A.P. denotes action pattern. S./ F. denote successful/ failure. 
The numbers in parentheses in the column headers indicate the total number of successful/ failure cases across four context configurations. 
For each stage and A.P., the table reports the number (ratio in \%) of \textit{successful cases that include}/ \textit{failure cases that miss} it. }
\label{tab:taxonomy_table}
\resizebox{\textwidth}{!}{%
\setlength{\tabcolsep}{2pt}%
\renewcommand{\arraystretch}{1.3}%
\begin{tabular}{l|l|llll|llllllllllll}
\hline
\multicolumn{1}{c|}{\multirow{2}{*}{\textbf{Stage}}}         & \multicolumn{1}{c|}{\multirow{2}{*}{\textbf{\begin{tabular}[c]{@{}c@{}}A.P.\end{tabular}}}} & \multicolumn{4}{c|}{\textbf{Hunk-level + Failure log}}                                                                                                                                                                                                     & \multicolumn{4}{c|}{\textbf{Method-level + Failure log}}                                                                                                                                                                                                             & \multicolumn{4}{c|}{\textbf{File-level + Failure log}}                                                                                                                                                                                                                  & \multicolumn{4}{c}{\textbf{\begin{tabular}[c]{@{}c@{}}File-level + Failure log\\+ Top-5 support\end{tabular}}}                                                                                                                                                                                         \\ \cline{3-18} 
\multicolumn{1}{c|}{}                                        & \multicolumn{1}{c|}{}                                              & \multicolumn{2}{c|}{\footnotesize\renewcommand{\arraystretch}{0.9}\textbf{\begin{tabular}[c]{@{}c@{}}\# S. cases include \\ the A.P. (234)\end{tabular}}}    & \multicolumn{2}{c|}{\footnotesize\renewcommand{\arraystretch}{0.9}\textbf{\begin{tabular}[c]{@{}c@{}}\# F. cases miss \\ the A.P. (266)\end{tabular}}} & \multicolumn{2}{c|}{\footnotesize\renewcommand{\arraystretch}{0.9}\textbf{\begin{tabular}[c]{@{}c@{}}\# S. cases include \\ the A.P. (229)\end{tabular}}}     & \multicolumn{2}{c|}{\footnotesize\renewcommand{\arraystretch}{0.9}\textbf{\begin{tabular}[c]{@{}c@{}}\# F. cases miss \\ the A.P. (271)\end{tabular}}}           & \multicolumn{2}{c|}{\footnotesize\renewcommand{\arraystretch}{0.9}\textbf{\begin{tabular}[c]{@{}c@{}}\# S. cases include \\ the A.P. (328)\end{tabular}}}         & \multicolumn{2}{c|}{\footnotesize\renewcommand{\arraystretch}{0.9}\textbf{\begin{tabular}[c]{@{}c@{}}\# F. cases miss \\ the A.P. (172)\end{tabular}}}            & \multicolumn{2}{c|}{\footnotesize\renewcommand{\arraystretch}{0.9}\textbf{\begin{tabular}[c]{@{}c@{}}\# S. cases include \\ the A.P. (334)\end{tabular}}}        & \multicolumn{2}{c}{\footnotesize\renewcommand{\arraystretch}{0.9}\textbf{\begin{tabular}[c]{@{}c@{}}\# F. cases miss \\ the A.P. (166)\end{tabular}}} \\ \hline
\multirow{3}{*}{\begin{tabular}[c]{@{}l@{}}\textbf{1}\end{tabular}} & \textit{A}                          & \multicolumn{1}{l|}{\multirow{3}{*}{\begin{tabular}[c]{@{}l@{}}230 \\ (98.3)\end{tabular}}} & \multicolumn{1}{l|}{184 (78.6)} & \multicolumn{1}{l|}{\multirow{3}{*}{\begin{tabular}[c]{@{}l@{}}99 \\ (37.2)\end{tabular}}}        & 52 (19.5)       & \multicolumn{1}{l|}{\multirow{3}{*}{\begin{tabular}[c]{@{}l@{}}224 \\ (97.8)\end{tabular}}} & \multicolumn{1}{l|}{173 (75.5)} & \multicolumn{1}{l|}{\multirow{3}{*}{\begin{tabular}[c]{@{}l@{}}96 \\ (35.4)\end{tabular}}} & \multicolumn{1}{l|}{51 (18.8)}  & \multicolumn{1}{l|}{\multirow{3}{*}{\begin{tabular}[c]{@{}l@{}}317 \\ (96.7)\end{tabular}}} & \multicolumn{1}{l|}{234 (71.3)}   & \multicolumn{1}{l|}{\multirow{3}{*}{\begin{tabular}[c]{@{}l@{}}70 \\ (40.7)\end{tabular}}}  & \multicolumn{1}{l|}{44 (25.6)}  & \multicolumn{1}{l|}{\multirow{3}{*}{\begin{tabular}[c]{@{}l@{}}323  \\ (96.7)\end{tabular}}} & \multicolumn{1}{l|}{216 (64.7)}   & \multicolumn{1}{l|}{\multirow{3}{*}{\begin{tabular}[c]{@{}l@{}}77 \\ (46.4)\end{tabular}}}                                                           & 54 (32.5)        \\ \cline{2-2} \cline{4-4} \cline{6-6} \cline{8-8} \cline{10-10} \cline{12-12} \cline{14-14} \cline{16-16} \cline{18-18} 
                                                             & \textit{B}                          & \multicolumn{1}{l|}{}                                                                         & \multicolumn{1}{l|}{203 (86.8)} & \multicolumn{1}{l|}{}                                                                              & 12 (4.51)       & \multicolumn{1}{l|}{}                                                                         & \multicolumn{1}{l|}{192 (83.8)} & \multicolumn{1}{l|}{}                                                                       & \multicolumn{1}{l|}{15 (5.54)}  & \multicolumn{1}{l|}{}                                                                         & \multicolumn{1}{l|}{287   (87.5)} & \multicolumn{1}{l|}{}                                                                        & \multicolumn{1}{l|}{17 (9.88)}  & \multicolumn{1}{l|}{}                                                                          & \multicolumn{1}{l|}{282   (84.4)} & \multicolumn{1}{l|}{}                                                                            & 17   (10.2)      \\ \cline{2-2} \cline{4-4} \cline{6-6} \cline{8-8} \cline{10-10} \cline{12-12} \cline{14-14} \cline{16-16} \cline{18-18} 
                                                             & \textit{C}                                  & \multicolumn{1}{l|}{}                                                                         & \multicolumn{1}{l|}{165 (70.5)} & \multicolumn{1}{l|}{}                                                                              & 50 (18.8)       & \multicolumn{1}{l|}{}                                                                         & \multicolumn{1}{l|}{147 (64.2)} & \multicolumn{1}{l|}{}                                                                       & \multicolumn{1}{l|}{49 (18.1)}  & \multicolumn{1}{l|}{}                                                                         & \multicolumn{1}{l|}{214   (65.2)} & \multicolumn{1}{l|}{}                                                                        & \multicolumn{1}{l|}{30 (17.4)}  & \multicolumn{1}{l|}{}                                                                          & \multicolumn{1}{l|}{234   (70.1)} & \multicolumn{1}{l|}{}                                                                            & 36   (21.7)      \\ \hline
\multirow{4}{*}{\begin{tabular}[c]{@{}l@{}}\textbf{2}\end{tabular}}        & \textit{D}                                 & \multicolumn{1}{l|}{\multirow{4}{*}{\begin{tabular}[c]{@{}l@{}}234 \\ (100)\end{tabular}}}  & \multicolumn{1}{l|}{233 (99.6)} & \multicolumn{1}{l|}{\multirow{4}{*}{\begin{tabular}[c]{@{}l@{}}217 \\ (81.6)\end{tabular}}}        & 0 (0.00)        & \multicolumn{1}{l|}{\multirow{4}{*}{\begin{tabular}[c]{@{}l@{}}229 \\ (100)\end{tabular}}}  & \multicolumn{1}{l|}{229 (100)}  & \multicolumn{1}{l|}{\multirow{4}{*}{\begin{tabular}[c]{@{}l@{}}226 \\ (83.4)\end{tabular}}} & \multicolumn{1}{l|}{1 (0.37)}   & \multicolumn{1}{l|}{\multirow{4}{*}{\begin{tabular}[c]{@{}l@{}}328 \\ (100)\end{tabular}}}  & \multicolumn{1}{l|}{328   (100)}  & \multicolumn{1}{l|}{\multirow{4}{*}{\begin{tabular}[c]{@{}l@{}}145 \\ (84.3)\end{tabular}}}  & \multicolumn{1}{l|}{4 (2.33)}   & \multicolumn{1}{l|}{\multirow{4}{*}{\begin{tabular}[c]{@{}l@{}}334 \\ (100)\end{tabular}}}   & \multicolumn{1}{l|}{333   (99.7)} & \multicolumn{1}{l|}{\multirow{4}{*}{\begin{tabular}[c]{@{}l@{}}143 \\ (86.1)\end{tabular}}}      & 2   (1.20)       \\ \cline{2-2} \cline{4-4} \cline{6-6} \cline{8-8} \cline{10-10} \cline{12-12} \cline{14-14} \cline{16-16} \cline{18-18} 
                                                             & \textit{E}                & \multicolumn{1}{l|}{}                                                                         & \multicolumn{1}{l|}{234 (100)}  & \multicolumn{1}{l|}{}                                                                              & 0 (0.00)        & \multicolumn{1}{l|}{}                                                                         & \multicolumn{1}{l|}{227 (99.1)} & \multicolumn{1}{l|}{}                                                                       & \multicolumn{1}{l|}{1 (0.37)}   & \multicolumn{1}{l|}{}                                                                         & \multicolumn{1}{l|}{327   (99.7)} & \multicolumn{1}{l|}{}                                                                        & \multicolumn{1}{l|}{4 (2.33)}   & \multicolumn{1}{l|}{}                                                                          & \multicolumn{1}{l|}{333   (99.7)} & \multicolumn{1}{l|}{}                                                                            & 3   (1.81)       \\ \cline{2-2} \cline{4-4} \cline{6-6} \cline{8-8} \cline{10-10} \cline{12-12} \cline{14-14} \cline{16-16} \cline{18-18} 
                                                             & \textit{F}                             & \multicolumn{1}{l|}{}                                                                         & \multicolumn{1}{l|}{132 (56.4)} & \multicolumn{1}{l|}{}                                                                              & 83 (31.2)       & \multicolumn{1}{l|}{}                                                                         & \multicolumn{1}{l|}{143 (62.4)} & \multicolumn{1}{l|}{}                                                                       & \multicolumn{1}{l|}{89 (32.8)}  & \multicolumn{1}{l|}{}                                                                         & \multicolumn{1}{l|}{200   (61.0)} & \multicolumn{1}{l|}{}                                                                        & \multicolumn{1}{l|}{71 (41.3)}  & \multicolumn{1}{l|}{}                                                                          & \multicolumn{1}{l|}{211   (63.2)} & \multicolumn{1}{l|}{}                                                                            & 75   (45.2)      \\ \cline{2-2} \cline{4-4} \cline{6-6} \cline{8-8} \cline{10-10} \cline{12-12} \cline{14-14} \cline{16-16} \cline{18-18} 
                                                             & \textit{G}                                 & \multicolumn{1}{l|}{}                                                                         & \multicolumn{1}{l|}{44 (18.8)}  & \multicolumn{1}{l|}{}                                                                              & 215 (80.8)      & \multicolumn{1}{l|}{}                                                                         & \multicolumn{1}{l|}{45 (19.7)}  & \multicolumn{1}{l|}{}                                                                       & \multicolumn{1}{l|}{226 (83.4)} & \multicolumn{1}{l|}{}                                                                         & \multicolumn{1}{l|}{68   (20.7)}  & \multicolumn{1}{l|}{}                                                                        & \multicolumn{1}{l|}{145 (84.3)} & \multicolumn{1}{l|}{}                                                                          & \multicolumn{1}{l|}{55   (16.5)}  & \multicolumn{1}{l|}{}                                                                            & 141   (84.9)     \\ \hline
\multirow{3}{*}{\begin{tabular}[c]{@{}l@{}}\textbf{3}\end{tabular}}           & \textit{H}                                 & \multicolumn{1}{l|}{\multirow{3}{*}{\begin{tabular}[c]{@{}l@{}}211 \\ (90.2)\end{tabular}}} & \multicolumn{1}{l|}{200 (85.5)} & \multicolumn{1}{l|}{\multirow{3}{*}{\begin{tabular}[c]{@{}l@{}}248 \\ (93.2)\end{tabular}}}       & 32 (12.0)       & \multicolumn{1}{l|}{\multirow{3}{*}{\begin{tabular}[c]{@{}l@{}}215 \\ (93.9)\end{tabular}}} & \multicolumn{1}{l|}{200 (87.3)} & \multicolumn{1}{l|}{\multirow{3}{*}{\begin{tabular}[c]{@{}l@{}}249 \\ (91.9)\end{tabular}}}                                            & \multicolumn{1}{l|}{22 (8.12)}  & \multicolumn{1}{l|}{\multirow{3}{*}{\begin{tabular}[c]{@{}l@{}}315 \\ (96.0)\end{tabular}}} & \multicolumn{1}{l|}{297   (90.6)} & \multicolumn{1}{l|}{\multirow{3}{*}{\begin{tabular}[c]{@{}l@{}}164 \\ (95.3)\end{tabular}}} & \multicolumn{1}{l|}{28 (16.3)}  & \multicolumn{1}{l|}{\multirow{3}{*}{\begin{tabular}[c]{@{}l@{}}322 \\ (96.4)\end{tabular}}}  & \multicolumn{1}{l|}{306   (91.6)} & \multicolumn{1}{l|}{\multirow{3}{*}{\begin{tabular}[c]{@{}l@{}}150 \\ (90.4)\end{tabular}}}     & 22   (13.3)      \\ \cline{2-2} \cline{4-4} \cline{6-6} \cline{8-8} \cline{10-10} \cline{12-12} \cline{14-14} \cline{16-16} \cline{18-18}
                                                             & \textit{I}                   & \multicolumn{1}{l|}{}                                                                         & \multicolumn{1}{l|}{38 (16.2)}  & \multicolumn{1}{l|}{}                                                                              & 227 (85.3)      & \multicolumn{1}{l|}{}                                                                         & \multicolumn{1}{l|}{33 (14.4)}  & \multicolumn{1}{l|}{}                                                                       & \multicolumn{1}{l|}{229 (84.5)} & \multicolumn{1}{l|}{}                                                                         & \multicolumn{1}{l|}{22   (6.71)}  & \multicolumn{1}{l|}{}                                                                        & \multicolumn{1}{l|}{153 (89.0)} & \multicolumn{1}{l|}{}                                                                          & \multicolumn{1}{l|}{34   (10.2)}  & \multicolumn{1}{l|}{}                                                                            & 141   (84.9)     \\ \cline{2-2} \cline{4-4} \cline{6-6} \cline{8-8} \cline{10-10} \cline{12-12} \cline{14-14} \cline{16-16} \cline{18-18} 
                                                             & \textit{J}                          & \multicolumn{1}{l|}{}                                                                         & \multicolumn{1}{l|}{106 (45.3)} & \multicolumn{1}{l|}{}                                                                              & 113 (42.5)      & \multicolumn{1}{l|}{}                                                                         & \multicolumn{1}{l|}{121 (52.8)} & \multicolumn{1}{l|}{}                                                                       & \multicolumn{1}{l|}{111 (41.0)} & \multicolumn{1}{l|}{}                                                                         & \multicolumn{1}{l|}{189   (57.6)} & \multicolumn{1}{l|}{}                                                                        & \multicolumn{1}{l|}{63 (36.6)}  & \multicolumn{1}{l|}{}                                                                          & \multicolumn{1}{l|}{170   (50.9)} & \multicolumn{1}{l|}{}                                                                            & 62   (37.3)      \\ \hline
\multirow{4}{*}{\begin{tabular}[c]{@{}l@{}}\textbf{4}\end{tabular}}         & \textit{K}                                     & \multicolumn{1}{l|}{\multirow{4}{*}{\begin{tabular}[c]{@{}l@{}}234 \\ (100)\end{tabular}}}  & \multicolumn{1}{l|}{228 (97.4)} & \multicolumn{1}{l|}{\multirow{4}{*}{\begin{tabular}[c]{@{}l@{}}255 \\ (95.9)\end{tabular}}}        & 2 (0.75)        & \multicolumn{1}{l|}{\multirow{4}{*}{\begin{tabular}[c]{@{}l@{}}229 \\ (100)\end{tabular}}}  & \multicolumn{1}{l|}{226 (98.7)} & \multicolumn{1}{l|}{\multirow{4}{*}{\begin{tabular}[c]{@{}l@{}}239 \\ (88.2)\end{tabular}}}                                            & \multicolumn{1}{l|}{2 (0.74)}   & \multicolumn{1}{l|}{\multirow{4}{*}{\begin{tabular}[c]{@{}l@{}}327 \\ (99.7)\end{tabular}}} & \multicolumn{1}{l|}{321 (97.9)}   & \multicolumn{1}{l|}{\multirow{4}{*}{\begin{tabular}[c]{@{}l@{}}158 \\ (91.9)\end{tabular}}}  & \multicolumn{1}{l|}{5 (2.91)}   & \multicolumn{1}{l|}{\multirow{4}{*}{\begin{tabular}[c]{@{}l@{}}333 \\ (99.7)\end{tabular}}}  & \multicolumn{1}{l|}{328 (98.2)}   & \multicolumn{1}{l|}{\multirow{4}{*}{\begin{tabular}[c]{@{}l@{}}151 \\ (91.0)\end{tabular}}}      & 3 (1.81)         \\ \cline{2-2} \cline{4-4} \cline{6-6} \cline{8-8} \cline{10-10} \cline{12-12} \cline{14-14} \cline{16-16} \cline{18-18} 
                                                             & \textit{L}                                          & \multicolumn{1}{l|}{}                                                                         & \multicolumn{1}{l|}{129 (55.1)} & \multicolumn{1}{l|}{}                                                                              & 143 (53.8)      & \multicolumn{1}{l|}{}                                                                         & \multicolumn{1}{l|}{108 (47.2)} & \multicolumn{1}{l|}{}                                                                       & \multicolumn{1}{l|}{130 (48.0)} & \multicolumn{1}{l|}{}                                                                         & \multicolumn{1}{l|}{146   (44.5)} & \multicolumn{1}{l|}{}                                                                        & \multicolumn{1}{l|}{84 (48.8)}  & \multicolumn{1}{l|}{}                                                                          & \multicolumn{1}{l|}{155   (46.4)} & \multicolumn{1}{l|}{}                                                                            & 77   (46.4)      \\ \cline{2-2} \cline{4-4} \cline{6-6} \cline{8-8} \cline{10-10} \cline{12-12} \cline{14-14} \cline{16-16} \cline{18-18} 
                                                             & \textit{M}                                     & \multicolumn{1}{l|}{}                                                                         & \multicolumn{1}{l|}{232 (99.1)} & \multicolumn{1}{l|}{}                                                                              & 3 (1.13)        & \multicolumn{1}{l|}{}                                                                         & \multicolumn{1}{l|}{224 (97.8)} & \multicolumn{1}{l|}{}                                                                       & \multicolumn{1}{l|}{8 (2.95)}   & \multicolumn{1}{l|}{}                                                                         & \multicolumn{1}{l|}{314   (95.7)} & \multicolumn{1}{l|}{}                                                                        & \multicolumn{1}{l|}{14 (8.14)}  & \multicolumn{1}{l|}{}                                                                          & \multicolumn{1}{l|}{324   (97.0)} & \multicolumn{1}{l|}{}                                                                            & 13   (7.83)      \\ \cline{2-2} \cline{4-4} \cline{6-6} \cline{8-8} \cline{10-10} \cline{12-12} \cline{14-14} \cline{16-16} \cline{18-18} 
                                                             & \textit{N}                                & \multicolumn{1}{l|}{}                                                                         & \multicolumn{1}{l|}{23 (9.83)}  & \multicolumn{1}{l|}{}                                                                              & 241 (90.6)      & \multicolumn{1}{l|}{}                                                                         & \multicolumn{1}{l|}{21 (9.17)}  & \multicolumn{1}{l|}{}                                                                       & \multicolumn{1}{l|}{228 (84.1)} & \multicolumn{1}{l|}{}                                                                         & \multicolumn{1}{l|}{34   (10.4)}  & \multicolumn{1}{l|}{}                                                                        & \multicolumn{1}{l|}{149 (86.6)} & \multicolumn{1}{l|}{}                                                                          & \multicolumn{1}{l|}{36   (10.8)}  & \multicolumn{1}{l|}{}                                                                            & 138   (83.1)     \\ \hline
\end{tabular}%
}
\vspace{-3mm}
\end{table*}

\textbf{Stage 1. Log-guided failure triage.}
The model begins by diagnosing test failure types and likely reasons from failure logs. Three types of actions are often observed in this stage:
\begin{enumerate*}[label=(\Alph*), start=1]
% \begin{itemize}[leftmargin=*,topsep=1pt,itemsep=1pt]
    \item {\textit{Failure type classification}}: Classify the failure into compilation error or runtime error.  
    \item {\textit{Broken element identification}}: Identify the broken symbols, invocations, assertions, or statements based on the failure log and broken test. 
    \item {\textit{Failure summarization}}: Analyze the failure's root cause and generate a concise summary before proceeding to repair. 
\end{enumerate*}
% \end{itemize}

\textbf{Stage 2. CUT delta analysis.}
The model then reasons about code symbols and program behavior changes from CUT diff, performing the following four typical actions:
\begin{enumerate*}[label=(\Alph*), start=4]
% \begin{itemize}[leftmargin=*,topsep=1pt,itemsep=1pt]
    \item {\textit{CUT diff summarization}}: Summarize symbol \eg{API} and behavior changes relevant to the test failure in the CUT before editing the test. 
    \item {\textit{Old-to-new element mapping construction}}: Construct a mapping from obsolete test elements to their replacements in the new version. 
    \item {\textit{Signature update detection}}: Detect constructor or method signature changes and identify missing/ changed arguments. 
    \item {\textit{New parameter handling}}: Identify newly required parameters and select appropriate values \eg{existing variables, null, default or mock values}. 
\end{enumerate*}
% \end{itemize}

\textbf{Stage 3. Repair planning.}
Next, the model plans concrete repair actions based on the previous failure diagnosis and CUT analysis before editing code, performing three typical actions:
\begin{enumerate*}[label=(\Alph*), start=8]
% \begin{itemize}[leftmargin=*,topsep=1pt,itemsep=1pt]
    \item {\textit{Stepwise decomposition}}: Break the repair into an ordered sequence of edits or plan steps. 
    \item {\textit{Referring to relevant usage examples}}: Refer to relevant code snippets in the codebase to use as replacement idioms for the updated API. 
    \item {\textit{Original test intent analysis}}: Infer the original test's verification goal to preserve it in the repair. 
\end{enumerate*}
% \end{itemize}

\textbf{Stage 4. Repair generation.}
Finally, the model generates the repaired test and verifies output correctness before responding with four typical actions:
\begin{enumerate*}[label=(\Alph*), start=11]
% \begin{itemize}[leftmargin=*,topsep=1pt,itemsep=1pt]
    \item {\textit{Test method update}}: Generate the repaired test method. 
    \item {\textit{Oracle update}}: Update expected values in test assertions to reflect the new behavior. 
    \item {\textit{Imports management}}: Add import statements only when needed and remove obsolete ones. 
    \item {\textit{Self-check (reflection)}}: Check the format, syntax, and semantic correctness of the test. 
\end{enumerate*}
% \end{itemize}

This taxonomy reveals that effective test repair by the LLM with reasoning mode enabled is a structured, multi-stage reasoning process that mirrors how an expert developer diagnoses a failure, understands the changed code, plans the fix, and verifies the result. 
Table~\ref{tab:taxonomy_table} presents statistics for the four stages and 14 action patterns across context configurations.  
We report the number and proportion of successful cases that include each stage or action pattern and the failed cases that miss each stage or action pattern. 
Overall, all four stages are observed in reasoning traces of over 90\% successful cases. 
Six action patterns \ie{B, D, E, H, K, M} appear in more than 70\% of successful cases for all context configurations. 
Meanwhile, action patterns A, C, and F appear in over 60\% of successful cases while being missed from about 20\%--30\% of failed cases. 
These observations suggest that the identified reasoning stages and action patterns are commonly involved in test repair by the LLMs with reasoning mode enabled; meanwhile, the patterns frequently missing from failed cases may serve as critical instructions to incentivize LLMs' success.

To assess the practical usefulness of the identified taxonomy, we further design the following two reasoning strategies for LLMs based on it:
\begin{itemize}[leftmargin=*,topsep=1pt,itemsep=1pt]
    \item \textbf{Optimized single-stage prompting} guides LLMs to repair tests in a single query, with the prompt structured to instruct the model through each reasoning stage and action patterns. 
    \item \textbf{Optimized multi-stage prompting} guides LLMs to decompose repair into four sequential queries following the divide-and-conquer intuition, one per stage, with each stage's output passed as context to the next.
\end{itemize}
The two strategies impose the taxonomy at different levels: single-stage prompting describes the process in the prompt at almost no extra cost, while multi-stage prompting enforces it through the pipeline at the price of more queries and longer context. 
The strategies aim to guide LLMs to perform test repair by following the validated effective process without requiring them to plan the reasoning steps themselves. 
The prompts are adapted from the SOTA LLM-based test update technique TestUpdater~\cite{DBLP:journals/corr/abs-2509-24419} and are available in our replication package in Section~\ref{sec:8}.

\begin{table}[t]
% \vspace{-1mm}
\setlength{\abovecaptionskip}{0pt}
\caption{Test repair performance of \textit{DeepSeek-V3.2} (\textit{abbr. DS-V3.2}) with/ without optimized prompting/ pipeline design strategies, and \textit{DS-V3.2} with reasoning mode enabled (\textit{DS-V3.2 (thinking)}).
}
% \vspace{-1mm}
\label{tab:RQ34}
\resizebox{\linewidth}{!}{%
\setlength{\tabcolsep}{2pt}%
\small%
\begin{tabular}{c|c|c|llll|llll|l}
\toprule
 & & & \multicolumn{4}{c|}{} & \multicolumn{4}{c|}{\textbf{Avg. Token Count per Case}} & \multicolumn{1}{c}{} \\ \cline{8-11}
 & & & \multicolumn{4}{c|}{\multirow{-2}{*}{\textbf{Performance}}} & \multicolumn{2}{c|}{\textit{Input}} & \multicolumn{1}{c|}{} & \multicolumn{1}{c|}{} & \multicolumn{1}{c}{} \\ \cline{4-9}
\multirow{-3}{*}{\textbf{Context Setup}} & \multirow{-3}{*}{\textbf{Model}} & \multirow{-3}{*}{\textbf{Strategy}} & \multicolumn{2}{c|}{\textit{exec}} & \multicolumn{2}{c|}{\textit{comp}} & \multicolumn{1}{c|}{\textit{cache miss}} & \multicolumn{1}{c|}{\textit{cache hit}} & \multicolumn{1}{c|}{\multirow{-2}{*}{\textit{Reasoning}}} & \multicolumn{1}{c|}{\multirow{-2}{*}{\textit{Output}}} & \multicolumn{1}{c}{\multirow{-3}{*}{\textbf{\begin{tabular}[c]{@{}c@{}}Avg. \\ Cost \\ (USD)\end{tabular}}}} \\ \hhline{*{12}{-}}
 & & \textit{baseline} & \multicolumn{1}{l|}{223} & \multicolumn{1}{l|}{0.446} & \multicolumn{1}{l|}{259} & 0.518 & \multicolumn{1}{l|}{4792} & \multicolumn{1}{l|}{\textbackslash{}} & \multicolumn{1}{l|}{\textbackslash{}} & 451 & 0.00344 \\ \hhline{~~*{10}{-}}
 & & \textit{single-stage} & \multicolumn{1}{l|}{226} & \multicolumn{1}{l|}{0.452} & \multicolumn{1}{l|}{270} & 0.540 & \multicolumn{1}{l|}{5233} & \multicolumn{1}{l|}{\textbackslash{}} & \multicolumn{1}{l|}{\textbackslash{}} & 433 & 0.00366 \\ \hhline{~~*{10}{-}}
 & \multirow{-3}{*}{\textit{DS-V3.2}} & \textit{multi-stage} & \multicolumn{1}{l|}{\textbf{244}} & \multicolumn{1}{l|}{\textbf{0.488}} & \multicolumn{1}{l|}{\textbf{292}} & \textbf{0.584} & \multicolumn{1}{l|}{5729} & \multicolumn{1}{l|}{12247} & \multicolumn{1}{l|}{\textbackslash{}} & 367 & 0.00468 \\ \hhline{~*{11}{-}}
\multirow{-4}{*}{\textit{\begin{tabular}[c]{@{}c@{}}Hunk-level +\\Failure log\end{tabular}}} & \textit{DS-V3.2 (thinking)} & \textit{baseline} & \multicolumn{1}{l|}{234} & \multicolumn{1}{l|}{0.468} & \multicolumn{1}{l|}{265} & 0.530 & \multicolumn{1}{l|}{4792} & \multicolumn{1}{l|}{\textbackslash{}} & \multicolumn{1}{l|}{2305} & 448 & 0.00731 \\ \hhline{*{12}{-}}
 & & \textit{baseline} & \multicolumn{1}{l|}{218} & \multicolumn{1}{l|}{0.436} & \multicolumn{1}{l|}{256} & 0.512 & \multicolumn{1}{l|}{5873} & \multicolumn{1}{l|}{\textbackslash{}} & \multicolumn{1}{l|}{\textbackslash{}} & 474 & 0.00408 \\ \hhline{~~*{10}{-}}
 & & \textit{single-stage} & \multicolumn{1}{l|}{233} & \multicolumn{1}{l|}{0.466} & \multicolumn{1}{l|}{261} & 0.522 & \multicolumn{1}{l|}{6314} & \multicolumn{1}{l|}{\textbackslash{}} & \multicolumn{1}{l|}{\textbackslash{}} & 435 & 0.00427 \\ \hhline{~~*{10}{-}}
 & \multirow{-3}{*}{\textit{DS-V3.2}} & \textit{multi-stage} & \multicolumn{1}{l|}{\textbf{260}} & \multicolumn{1}{l|}{\textbf{0.520}} & \multicolumn{1}{l|}{\textbf{306}} & \textbf{0.612} & \multicolumn{1}{l|}{6818} & \multicolumn{1}{l|}{14424} & \multicolumn{1}{l|}{\textbackslash{}} & 355 & 0.00542 \\ \hhline{~*{11}{-}}
\multirow{-4}{*}{\textit{\begin{tabular}[c]{@{}c@{}}Method-level\\+ Failure log\end{tabular}}} & \textit{DS-V3.2 (thinking)} & \textit{baseline} & \multicolumn{1}{l|}{229} & \multicolumn{1}{l|}{0.458} & \multicolumn{1}{l|}{260} & 0.520 & \multicolumn{1}{l|}{5872} & \multicolumn{1}{l|}{\textbackslash{}} & \multicolumn{1}{l|}{2368} & 471 & 0.00806 \\ \hhline{*{12}{-}}
 & & \textit{baseline} & \multicolumn{1}{l|}{304} & \multicolumn{1}{l|}{0.608} & \multicolumn{1}{l|}{354} & 0.708 & \multicolumn{1}{l|}{14875} & \multicolumn{1}{l|}{\textbackslash{}} & \multicolumn{1}{l|}{\textbackslash{}} & 452 & 0.00909 \\ \hhline{~~*{10}{-}}
 & & \textit{single-stage} & \multicolumn{1}{l|}{307} & \multicolumn{1}{l|}{0.614} & \multicolumn{1}{l|}{\textbf{364}} & \textbf{0.728} & \multicolumn{1}{l|}{15316} & \multicolumn{1}{l|}{\textbackslash{}} & \multicolumn{1}{l|}{\textbackslash{}} & 435 & 0.00931 \\ \hhline{~~*{10}{-}}
 & \multirow{-3}{*}{\textit{DS-V3.2}} & \textit{multi-stage} & \multicolumn{1}{l|}{305} & \multicolumn{1}{l|}{0.610} & \multicolumn{1}{l|}{360} & 0.720 & \multicolumn{1}{l|}{15445} & \multicolumn{1}{l|}{31366} & \multicolumn{1}{l|}{\textbackslash{}} & 353 & 0.01144 \\ \hhline{~*{11}{-}}
\multirow{-4}{*}{\textit{\begin{tabular}[c]{@{}c@{}}File-level +\\Failure log\end{tabular}}} & \textit{DS-V3.2 (thinking)} & \textit{baseline} & \multicolumn{1}{l|}{\textbf{328}} & \multicolumn{1}{l|}{\textbf{0.656}} & \multicolumn{1}{l|}{357} & 0.714 & \multicolumn{1}{l|}{14543} & \multicolumn{1}{l|}{\textbackslash{}} & \multicolumn{1}{l|}{2298} & 443 & 0.01275 \\ \hhline{*{12}{-}}
 & & \textit{baseline} & \multicolumn{1}{l|}{311} & \multicolumn{1}{l|}{0.622} & \multicolumn{1}{l|}{357} & 0.714 & \multicolumn{1}{l|}{15654} & \multicolumn{1}{l|}{\textbackslash{}} & \multicolumn{1}{l|}{\textbackslash{}} & 447 & 0.00952 \\ \hhline{~~*{10}{-}}
 & & \textit{single-stage} & \multicolumn{1}{l|}{323} & \multicolumn{1}{l|}{0.646} & \multicolumn{1}{l|}{387} & 0.774 & \multicolumn{1}{l|}{16095} & \multicolumn{1}{l|}{\textbackslash{}} & \multicolumn{1}{l|}{\textbackslash{}} & 432 & 0.00974 \\ \hhline{~~*{10}{-}}
 & \multirow{-3}{*}{\textit{DS-V3.2}} & \textit{multi-stage} & \multicolumn{1}{l|}{325} & \multicolumn{1}{l|}{0.650} & \multicolumn{1}{l|}{\textbf{389}} & \textbf{0.778} & \multicolumn{1}{l|}{15518} & \multicolumn{1}{l|}{31547} & \multicolumn{1}{l|}{\textbackslash{}} & 387 & 0.01155 \\ \hhline{~*{11}{-}}
\multirow{-4}{*}{\textit{\begin{tabular}[c]{@{}c@{}}File-level +\\Failure log +\\Top-5 support\end{tabular}}} & \textit{DS-V3.2 (thinking)} & \textit{baseline} & \multicolumn{1}{l|}{\textbf{334}} & \multicolumn{1}{l|}{\textbf{0.668}} & \multicolumn{1}{l|}{374} & 0.748 & \multicolumn{1}{l|}{15341} & \multicolumn{1}{l|}{\textbackslash{}} & \multicolumn{1}{l|}{2372} & 471 & 0.01337 \\ \bottomrule
\end{tabular}%
}
\vspace{-5mm}
\end{table}

We first apply these strategies to \textit{DeepSeek-V3.2} with reasoning mode disabled. 
Surprisingly, \textbf{\textit{our optimized prompting strategies achieve comparable or superior performance to reasoning mode at lower cost}}. 
As shown in Table~\ref{tab:RQ34}, both strategies (rows of \textit{DS-V3.2} with \textit{single-/multi-stage}) consistently improve \textit{DeepSeek-V3.2} over the baseline prompting (rows of \textit{DS-V3.2} with \textit{baseline}) across all four context configurations. 
Moreover, the multi-stage optimization strategy can achieve performance comparable to or even better than \textit{DeepSeek-V3.2} with reasoning mode enabled \eg{it achieves higher \textit{exec} with hunk- and method-level CUT changes and higher \textit{comp} under all four context configurations}. 
In addition, both strategies cost less than the original reasoning mode, since they produce no lengthy reasoning tokens and most of the additional input of the multi-stage strategy hits the cache. 
This suggests that the summarized effective reasoning process avoids unnecessary or even incorrect reasoning steps and paths, thus improving cost-effectiveness. 

To evaluate generalizability, we further apply both strategies to the other two LLMs (Table~\ref{tab:RQ34_others}).
We further observe that \textbf{\textit{the single-stage strategy generalizes well to all other LLMs}}. 
The single-stage strategy improves both \textit{exec} and \textit{comp} of both \textit{GPT-5.4} and \textit{Qwen3-Coder} across all four context configurations, with marginal additional cost (at most 5.7\%) \eg{with Method-level\,+\,Failure log, it relatively improves the \textit{exec} of \textit{GPT-5.4} by 11.0\% and that of \textit{Qwen3-Coder} by 10.6\%}. 
The multi-stage strategy also benefits \textit{GPT-5.4} under all four context configurations, but slightly decreases \textit{Qwen3-Coder}'s repair performance. 
A possible reason is that the four-stage pipeline accumulates intermediate context, to which \textit{Qwen3-Coder} appears more sensitive, as observed in RQ1. 
Overall, single-stage prompting is a robust default with consistent gains
at marginal cost, while multi-stage prompting can bring larger gains for certain combination of model and context but is costlier and model-dependent. 
These results suggest that the reasoning patterns distilled from \textit{DeepSeek-V3.2} can benefit LLMs beyond it.
\begin{table}[t]
% \vspace{-1mm}
\setlength{\abovecaptionskip}{0pt}
\caption{Test repair performance of \textit{Qwen3-Coder} and \textit{GPT-5.4} with/ without optimized prompting/ pipeline design strategies.
}
% \vspace{-1mm}
\label{tab:RQ34_others}
\resizebox{\linewidth}{!}{%
\setlength{\tabcolsep}{2pt}%
\small%
\begin{tabular}{c|c|c|llll|llll|l}
\toprule
 & & & \multicolumn{4}{c|}{} & \multicolumn{4}{c|}{\textbf{Avg. Token Count per Case}} & \multicolumn{1}{c}{} \\ \cline{8-11}
 & & & \multicolumn{4}{c|}{\multirow{-2}{*}{\textbf{Performance}}} & \multicolumn{2}{c|}{\textit{Input}} & \multicolumn{1}{c|}{} & \multicolumn{1}{c|}{} & \multicolumn{1}{c}{} \\ \cline{4-9}
\multirow{-3}{*}{\textbf{Context Setup}} & \multirow{-3}{*}{\textbf{Model}} & \multirow{-3}{*}{\textbf{Strategy}} & \multicolumn{2}{c|}{\textit{exec}} & \multicolumn{2}{c|}{\textit{comp}} & \multicolumn{1}{c|}{\textit{cache miss}} & \multicolumn{1}{c|}{\textit{cache hit}} & \multicolumn{1}{c|}{\multirow{-2}{*}{\textit{Reasoning}}} & \multicolumn{1}{c|}{\multirow{-2}{*}{\textit{Output}}} & \multicolumn{1}{c}{\multirow{-3}{*}{\textbf{\begin{tabular}[c]{@{}c@{}}Avg. \\ Cost \\ (USD)\end{tabular}}}} \\ \hhline{*{12}{-}}
 & & \textit{baseline} & \multicolumn{1}{l|}{245} & \multicolumn{1}{l|}{0.490} & \multicolumn{1}{l|}{285} & 0.570 & \multicolumn{1}{l|}{4705} & \multicolumn{1}{l|}{\textbackslash{}} & \multicolumn{1}{l|}{\textbackslash{}} & 401 & 0.00671 \\ \hhline{~~*{10}{-}}
 & & \textit{single-stage} & \multicolumn{1}{l|}{\textbf{266}} & \multicolumn{1}{l|}{\textbf{0.532}} & \multicolumn{1}{l|}{\textbf{311}} & \textbf{0.622} & \multicolumn{1}{l|}{5130} & \multicolumn{1}{l|}{\textbackslash{}} & \multicolumn{1}{l|}{\textbackslash{}} & 393 & 0.00709 \\ \hhline{~~*{10}{-}}
 & \multirow{-3}{*}{\textit{Qwen3-Coder}} & \textit{multi-stage} & \multicolumn{1}{l|}{229} & \multicolumn{1}{l|}{0.458} & \multicolumn{1}{l|}{285} & 0.570 & \multicolumn{1}{l|}{5610} & \multicolumn{1}{l|}{12202} & \multicolumn{1}{l|}{\textbackslash{}} & 436 & 0.00901 \\ \hhline{~*{11}{-}}
 & & \textit{baseline} & \multicolumn{1}{l|}{275} & \multicolumn{1}{l|}{0.550} & \multicolumn{1}{l|}{304} & 0.608 & \multicolumn{1}{l|}{4402} & \multicolumn{1}{l|}{\textbackslash{}} & \multicolumn{1}{l|}{\textbackslash{}} & 403 & 0.01704 \\ \hhline{~~*{10}{-}}
 & & \textit{single-stage} & \multicolumn{1}{l|}{\textbf{296}} & \multicolumn{1}{l|}{\textbf{0.592}} & \multicolumn{1}{l|}{\textbf{328}} & \textbf{0.656} & \multicolumn{1}{l|}{4826} & \multicolumn{1}{l|}{\textbackslash{}} & \multicolumn{1}{l|}{\textbackslash{}} & 386 & 0.01785 \\ \hhline{~~*{10}{-}}
\multirow{-6}{*}{\textit{\begin{tabular}[c]{@{}c@{}}Hunk-level +\\Failure log\end{tabular}}} & \multirow{-3}{*}{\textit{GPT-5.4}} & \textit{multi-stage} & \multicolumn{1}{l|}{278} & \multicolumn{1}{l|}{0.556} & \multicolumn{1}{l|}{318} & 0.636 & \multicolumn{1}{l|}{5303} & \multicolumn{1}{l|}{11267} & \multicolumn{1}{l|}{\textbackslash{}} & 470 & 0.02312 \\ \hhline{*{12}{-}}
 & & \textit{baseline} & \multicolumn{1}{l|}{273} & \multicolumn{1}{l|}{0.546} & \multicolumn{1}{l|}{308} & 0.616 & \multicolumn{1}{l|}{5756} & \multicolumn{1}{l|}{\textbackslash{}} & \multicolumn{1}{l|}{\textbackslash{}} & 400 & 0.00775 \\ \hhline{~~*{10}{-}}
 & & \textit{single-stage} & \multicolumn{1}{l|}{\textbf{302}} & \multicolumn{1}{l|}{\textbf{0.604}} & \multicolumn{1}{l|}{\textbf{342}} & \textbf{0.684} & \multicolumn{1}{l|}{6181} & \multicolumn{1}{l|}{\textbackslash{}} & \multicolumn{1}{l|}{\textbackslash{}} & 390 & 0.00813 \\ \hhline{~~*{10}{-}}
 & \multirow{-3}{*}{\textit{Qwen3-Coder}} & \textit{multi-stage} & \multicolumn{1}{l|}{250} & \multicolumn{1}{l|}{0.500} & \multicolumn{1}{l|}{304} & 0.608 & \multicolumn{1}{l|}{6609} & \multicolumn{1}{l|}{14188} & \multicolumn{1}{l|}{\textbackslash{}} & 429 & 0.01017 \\ \hhline{~*{11}{-}}
 & & \textit{baseline} & \multicolumn{1}{l|}{283} & \multicolumn{1}{l|}{0.566} & \multicolumn{1}{l|}{316} & 0.632 & \multicolumn{1}{l|}{5461} & \multicolumn{1}{l|}{\textbackslash{}} & \multicolumn{1}{l|}{\textbackslash{}} & 408 & 0.01977 \\ \hhline{~~*{10}{-}}
 & & \textit{single-stage} & \multicolumn{1}{l|}{\textbf{314}} & \multicolumn{1}{l|}{\textbf{0.628}} & \multicolumn{1}{l|}{\textbf{344}} & \textbf{0.688} & \multicolumn{1}{l|}{5885} & \multicolumn{1}{l|}{\textbackslash{}} & \multicolumn{1}{l|}{\textbackslash{}} & 386 & 0.02051 \\ \hhline{~~*{10}{-}}
\multirow{-6}{*}{\textit{\begin{tabular}[c]{@{}c@{}}Method-level\\+ Failure log\end{tabular}}} & \multirow{-3}{*}{\textit{GPT-5.4}} & \textit{multi-stage} & \multicolumn{1}{l|}{301} & \multicolumn{1}{l|}{0.602} & \multicolumn{1}{l|}{336} & 0.672 & \multicolumn{1}{l|}{6362} & \multicolumn{1}{l|}{13385} & \multicolumn{1}{l|}{\textbackslash{}} & 466 & 0.02625 \\ \hhline{*{12}{-}}
 & & \textit{baseline} & \multicolumn{1}{l|}{293} & \multicolumn{1}{l|}{0.586} & \multicolumn{1}{l|}{351} & 0.702 & \multicolumn{1}{l|}{14100} & \multicolumn{1}{l|}{\textbackslash{}} & \multicolumn{1}{l|}{\textbackslash{}} & 397 & 0.01608 \\ \hhline{~~*{10}{-}}
 & & \textit{single-stage} & \multicolumn{1}{l|}{\textbf{294}} & \multicolumn{1}{l|}{\textbf{0.588}} & \multicolumn{1}{l|}{\textbf{357}} & \textbf{0.714} & \multicolumn{1}{l|}{14525} & \multicolumn{1}{l|}{\textbackslash{}} & \multicolumn{1}{l|}{\textbackslash{}} & 387 & 0.01646 \\ \hhline{~~*{10}{-}}
 & \multirow{-3}{*}{\textit{Qwen3-Coder}} & \textit{multi-stage} & \multicolumn{1}{l|}{279} & \multicolumn{1}{l|}{0.558} & \multicolumn{1}{l|}{351} & 0.702 & \multicolumn{1}{l|}{14507} & \multicolumn{1}{l|}{29592} & \multicolumn{1}{l|}{\textbackslash{}} & 436 & 0.01965 \\ \hhline{~*{11}{-}}
 & & \textit{baseline} & \multicolumn{1}{l|}{321} & \multicolumn{1}{l|}{0.642} & \multicolumn{1}{l|}{364} & 0.728 & \multicolumn{1}{l|}{13876} & \multicolumn{1}{l|}{\textbackslash{}} & \multicolumn{1}{l|}{\textbackslash{}} & 420 & 0.04099 \\ \hhline{~~*{10}{-}}
 & & \textit{single-stage} & \multicolumn{1}{l|}{339} & \multicolumn{1}{l|}{0.678} & \multicolumn{1}{l|}{390} & 0.780 & \multicolumn{1}{l|}{14300} & \multicolumn{1}{l|}{\textbackslash{}} & \multicolumn{1}{l|}{\textbackslash{}} & 387 & 0.04156 \\ \hhline{~~*{10}{-}}
\multirow{-6}{*}{\textit{\begin{tabular}[c]{@{}c@{}}File-level +\\Failure log\end{tabular}}} & \multirow{-3}{*}{\textit{GPT-5.4}} & \textit{multi-stage} & \multicolumn{1}{l|}{\textbf{353}} & \multicolumn{1}{l|}{\textbf{0.706}} & \multicolumn{1}{l|}{\textbf{405}} & \textbf{0.810} & \multicolumn{1}{l|}{14777} & \multicolumn{1}{l|}{30214} & \multicolumn{1}{l|}{\textbackslash{}} & 466 & 0.05148 \\ \hhline{*{12}{-}}
 & & \textit{baseline} & \multicolumn{1}{l|}{301} & \multicolumn{1}{l|}{0.602} & \multicolumn{1}{l|}{359} & 0.718 & \multicolumn{1}{l|}{14823} & \multicolumn{1}{l|}{\textbackslash{}} & \multicolumn{1}{l|}{\textbackslash{}} & 396 & 0.01680 \\ \hhline{~~*{10}{-}}
 & & \textit{single-stage} & \multicolumn{1}{l|}{\textbf{317}} & \multicolumn{1}{l|}{\textbf{0.634}} & \multicolumn{1}{l|}{\textbf{379}} & \textbf{0.758} & \multicolumn{1}{l|}{15248} & \multicolumn{1}{l|}{\textbackslash{}} & \multicolumn{1}{l|}{\textbackslash{}} & 387 & 0.01718 \\ \hhline{~~*{10}{-}}
 & \multirow{-3}{*}{\textit{Qwen3-Coder}} & \textit{multi-stage} & \multicolumn{1}{l|}{284} & \multicolumn{1}{l|}{0.568} & \multicolumn{1}{l|}{351} & 0.702 & \multicolumn{1}{l|}{14672} & \multicolumn{1}{l|}{29953} & \multicolumn{1}{l|}{\textbackslash{}} & 447 & 0.01990 \\ \hhline{~*{11}{-}}
 & & \textit{baseline} & \multicolumn{1}{l|}{345} & \multicolumn{1}{l|}{0.690} & \multicolumn{1}{l|}{391} & 0.782 & \multicolumn{1}{l|}{14610} & \multicolumn{1}{l|}{\textbackslash{}} & \multicolumn{1}{l|}{\textbackslash{}} & 416 & 0.04277 \\ \hhline{~~*{10}{-}}
 & & \textit{single-stage} & \multicolumn{1}{l|}{351} & \multicolumn{1}{l|}{0.702} & \multicolumn{1}{l|}{400} & 0.800 & \multicolumn{1}{l|}{15034} & \multicolumn{1}{l|}{\textbackslash{}} & \multicolumn{1}{l|}{\textbackslash{}} & 388 & 0.04340 \\ \hhline{~~*{10}{-}}
\multirow{-6}{*}{\textit{\begin{tabular}[c]{@{}c@{}}File-level +\\Failure log +\\Top-5 support\end{tabular}}} & \multirow{-3}{*}{\textit{GPT-5.4}} & \textit{multi-stage} & \multicolumn{1}{l|}{\textbf{360}} & \multicolumn{1}{l|}{\textbf{0.720}} & \multicolumn{1}{l|}{\textbf{415}} & \textbf{0.830} & \multicolumn{1}{l|}{14803} & \multicolumn{1}{l|}{30292} & \multicolumn{1}{l|}{\textbackslash{}} & 517 & 0.05233 \\ \bottomrule
\end{tabular}%
}
\vspace{-5mm}
\end{table}

\begin{tcolorbox}[boxsep=2pt,left=2pt,right=2pt,top=2pt,bottom=2pt,width=\linewidth,colback=white!90!black,boxrule=0pt, colbacktitle=white!,toptitle=2pt,bottomtitle=1pt,opacitybacktitle=0]
\textbf{Answer to RQ3.}
Explicit reasoning traces improve model performance across all context settings. 
Reasoning traces from successful repairs reveal a taxonomy of a four-stage process with 14 common action patterns and proves effective for guiding LLMs to repair tests cost-effectively. 
\end{tcolorbox}

\subsection{RQ4. Performance across Scenario Complexity}
\begin{figure}[t!]
% \vspace{-3mm}
\centering
\renewcommand{\thesubfigure}{%
  \ifcase\value{subfigure}%
  \or a1%
  \or a2%
  \or a3%
  \or b1%
  \or b2%
  \or b3%
  \fi}

\begin{minipage}[t]{0.32\linewidth}
    \centering
    \includegraphics[width=\linewidth]{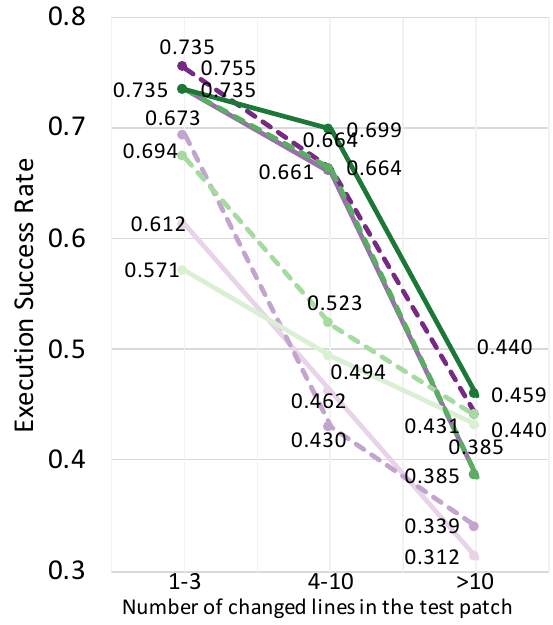}
    \Description{Line chart of execution success rate as the degree of test patch adaptation, measured in changed code lines, increases.}
    \subcaption{Perf. change with test patch adaptation degree by code lines (\textit{DS-V3.2}).}
    \label{fig:RQ5_a1}
\end{minipage}
% \vspace{1mm}
\hfill
\begin{minipage}[t]{0.32\linewidth}
    \centering
    \includegraphics[width=\linewidth]{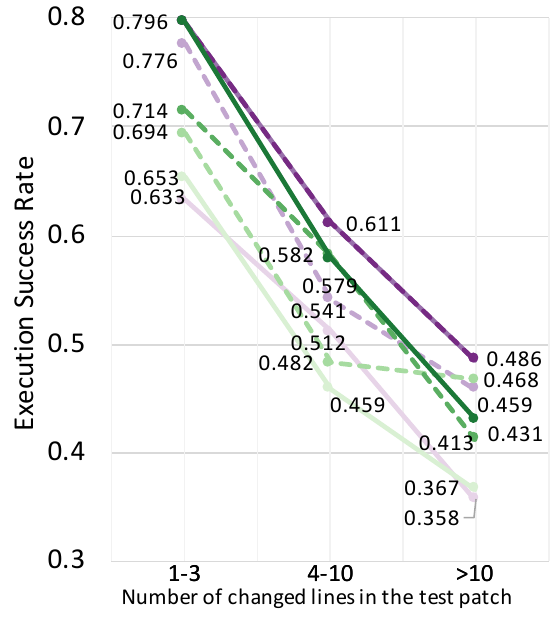}
    \Description{Line chart of execution success rate as the degree of test patch adaptation, measured in changed code hunks, increases.}
    \subcaption{Perf. change with test patch adaptation degree by code lines (\textit{Qwen3-Coder}).}
    \label{fig:RQ5_a2}
\end{minipage}
% \vspace{1mm}
\hfill
\begin{minipage}[t]{0.32\linewidth}
    \centering
    \includegraphics[width=\linewidth]{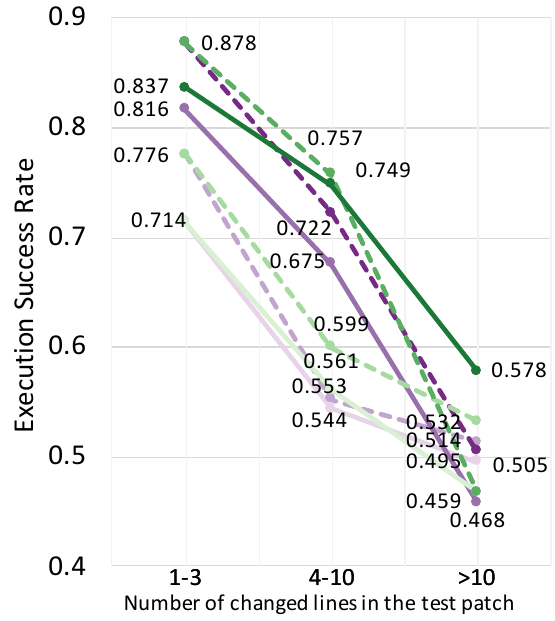}
    \Description{Line chart of execution success rate as the scale of changes to the code under test, measured in changed code lines, increases.}
    \subcaption{Perf. change with test patch adaptation degree by code lines (\textit{GPT-5.4}).}
    \label{fig:RQ5_a3}
\end{minipage}
% \vspace{1mm}
\hfill

\begin{minipage}[t]{0.32\linewidth}
    \centering
    \includegraphics[width=\linewidth]{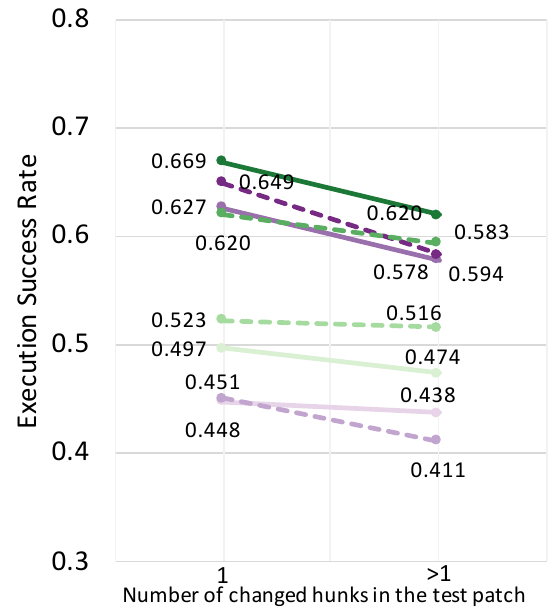}
    \Description{Line chart of execution success rate as the scale of changes to the code under test, measured in changed methods, increases.}
    \subcaption{Perf. change with test patch adaptation degree by code hunks (\textit{DS-V3.2}).}
    \label{fig:RQ5_b1}
\end{minipage}
\vspace{0.5mm}
\begin{minipage}[t]{0.32\linewidth}
    \centering
    \includegraphics[width=\linewidth]{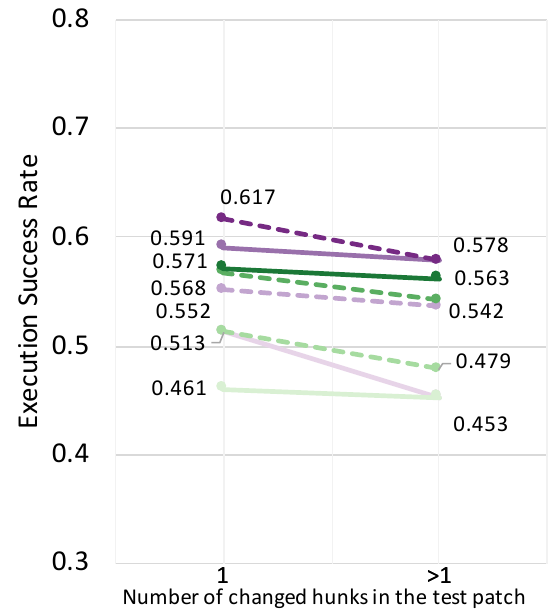}
    \Description{Line chart of execution success rate as the scale of changes to the code under test, measured in changed methods, increases.}
    \subcaption{Perf. change with test patch adaptation degree by code hunks (\textit{Qwen3-Coder}).}
    \label{fig:RQ5_b2}
\end{minipage}
\vspace{0.5mm}
\begin{minipage}[t]{0.32\linewidth}
    \centering
    \includegraphics[width=\linewidth]{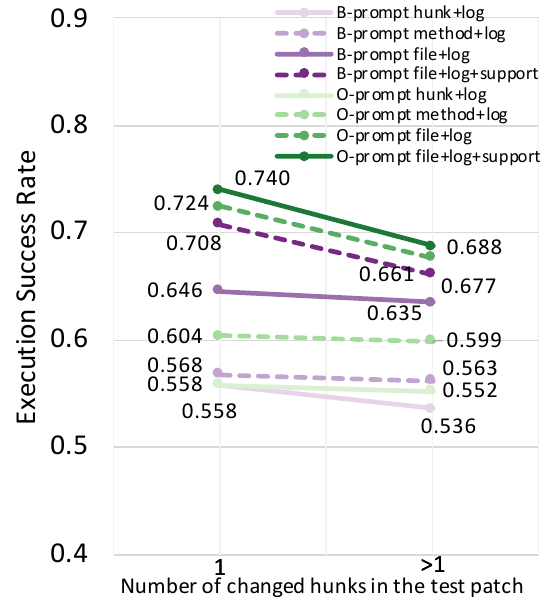}
    \Description{Line chart of execution success rate as the scale of changes to the code under test, measured in changed methods, increases.}
    \subcaption{Perf. change with test patch adaptation degree by code hunks (\textit{GPT-5.4}).}
    \label{fig:RQ5_b3}
\end{minipage}
\vspace{0.5mm}

\caption{Execution success rate (exec) of tests repaired by different LLMs with baseline prompting (\textit{abbr.}, B-prompt) and optimized multi-stage prompting (\textit{abbr.}, O-prompt) under different context configurations (Hunk-/Method-/File-level\,+\,Failure log\,+\,Top-5 Support) across scenario complexity levels measured by test patch adaptation degree.
The complexity is measured by test changed lines in (a1)-(a3) and by test changed hunks in (b1)-(b3).}
\label{fig:RQ5_test}
\vspace{-7mm}
\end{figure}

\begin{figure}[t!]
% \vspace{-3mm}
\centering
\renewcommand{\thesubfigure}{%
  \ifcase\value{subfigure}%
  \or c1%
  \or c2%
  \or c3%
  \or d1%
  \or d2%
  \or d3%
  \fi}

\begin{minipage}[t]{0.32\linewidth}
    \centering
    \includegraphics[width=\linewidth]{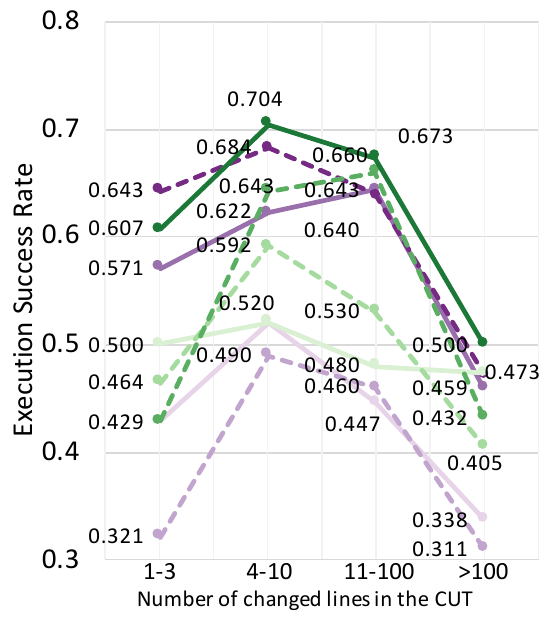}
    \Description{Line chart of execution success rate as the degree of test patch adaptation, measured in changed code lines, increases.}
    \subcaption{Perf. change with CUT change scale by code lines (\textit{DS-V3.2}).}
    \label{fig:RQ5_c1}
\end{minipage}
\hfill
\begin{minipage}[t]{0.32\linewidth}
    \centering
    \includegraphics[width=\linewidth]{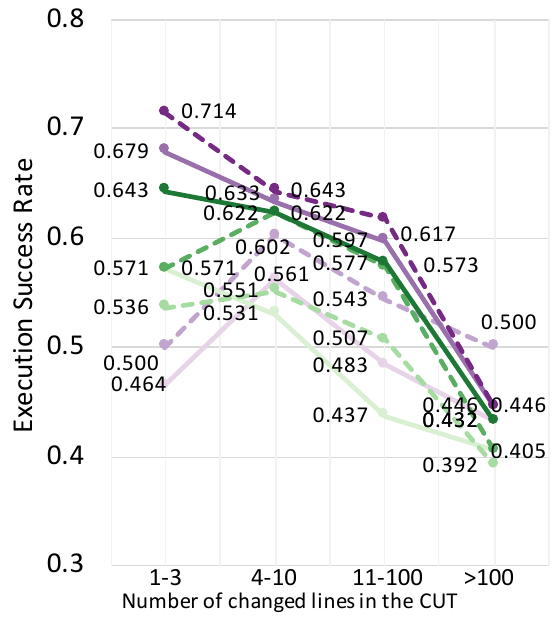}
    \Description{Line chart of execution success rate as the degree of test patch adaptation, measured in changed code hunks, increases.}
    \subcaption{Perf. change with CUT change scale by code lines (\textit{Qwen3-Coder}).}
    \label{fig:RQ5_c2}
\end{minipage}
\hfill
\begin{minipage}[t]{0.32\linewidth}
    \centering
    \includegraphics[width=\linewidth]{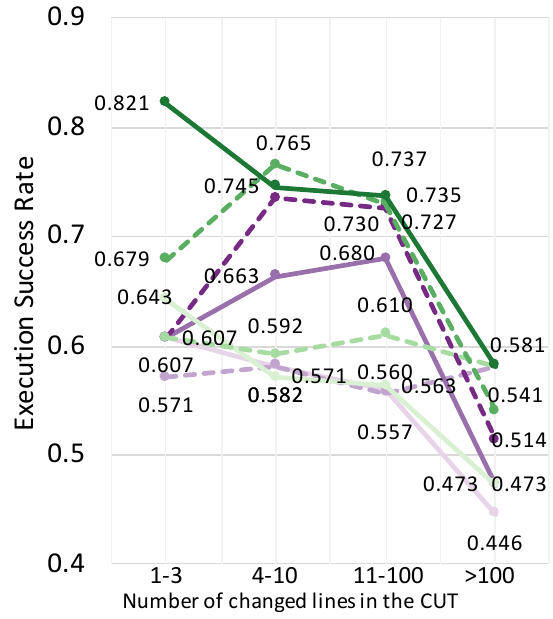}
    \Description{Line chart of execution success rate as the scale of changes to the code under test, measured in changed code lines, increases.}
    \subcaption{Perf. change with CUT change scale by code lines (\textit{GPT-5.4}).}
    \label{fig:RQ5_c3}
\end{minipage}
\hfill

\begin{minipage}[t]{0.32\linewidth}
    \centering
    \includegraphics[width=\linewidth]{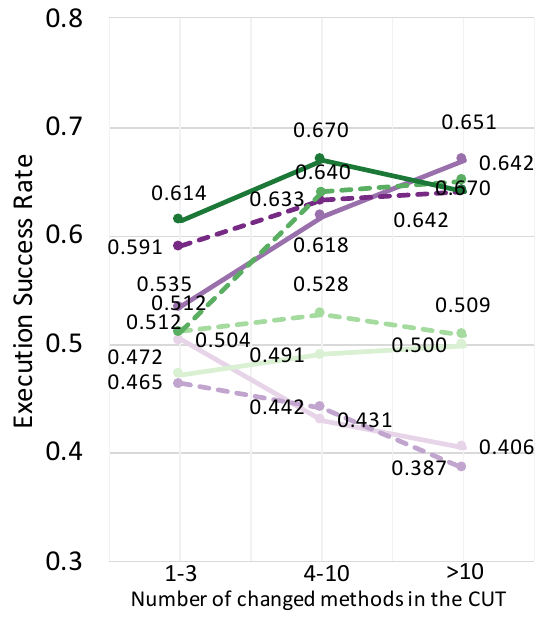}
    \Description{Line chart of execution success rate as the scale of changes to the code under test, measured in changed methods, increases.}
    \subcaption{Perf. change with CUT change scale by methods (\textit{DS-V3.2}).}
    \label{fig:RQ5_d1}
\end{minipage}
\vspace{0.5mm}
\begin{minipage}[t]{0.32\linewidth}
    \centering
    \includegraphics[width=\linewidth]{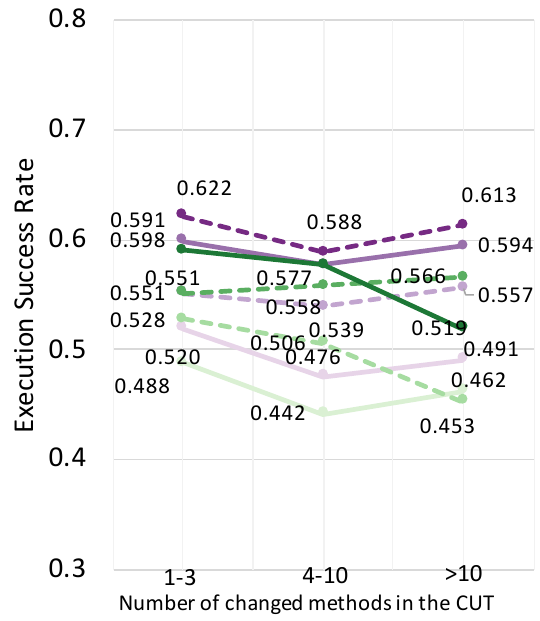}
    \Description{Line chart of execution success rate as the scale of changes to the code under test, measured in changed methods, increases.}
    \subcaption{Perf. change with CUT change scale by methods (\textit{Qwen3-Coder}).}
    \label{fig:RQ5_d2}
\end{minipage}
\vspace{0.5mm}
\begin{minipage}[t]{0.32\linewidth}
    \centering
    \includegraphics[width=\linewidth]{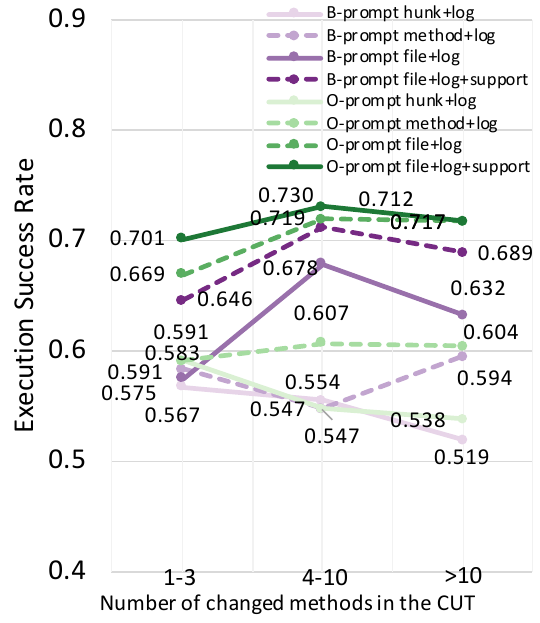}
    \Description{Line chart of execution success rate as the scale of changes to the code under test, measured in changed methods, increases.}
    \subcaption{Perf. change with CUT change scale by methods (\textit{GPT-5.4}).}
    \label{fig:RQ5_d3}
\end{minipage}

\vspace{0.5mm}

\caption{Execution success rate (exec) of tests repaired by different LLMs with baseline prompting (\textit{abbr.}, B-prompt) and optimized multi-stage prompting (\textit{abbr.}, O-prompt) under different context configurations (Hunk-/Method-/File-level\,+\,Failure log\,+\,Top-5 Support) across scenario complexity levels measured by CUT change scale.
The complexity is measured by CUT changed lines in (c1)-(c3) and by CUT changed methods in (d1)-(d3).}

\label{fig:RQ5_cut}
\vspace{-5mm}
\end{figure}

In RQ4, we study how repair performance changes as the complexity of source--test co-evolution scenarios increases. 
As repair difficulty cannot be measured directly, we approximate the complexity of a co-evolution scenario by its change scale on both the test and CUT sides, since larger changes generally require LLMs to understand more modifications and produce more edits. 
Specifically, we measure the complexity along two dimensions. 
\begin{itemize}[leftmargin=*,topsep=1pt,itemsep=1pt]
     \item \textbf{Test patch adaptation degree}, which is quantified by the number of changed code lines and hunks in the broken test's golden patch, reflecting the adaptation effort needed to align the broken test with the new CUT behavior. 
     \item \textbf{CUT change scale}, which is quantified by the number of changed lines and methods in the CUT, as larger-scale changes are more likely to introduce complex, interdependent co-evolution scenarios that require more sophisticated understanding to repair the broken test. 
\end{itemize}
We categorize repair cases into different complexity levels and analyze repair performance within each category. 
Figures~\ref{fig:RQ5_test} and~\ref{fig:RQ5_cut} show the repair performance of three LLMs \ie{\textit{DeepSeek-V3.2}, \textit{Qwen3-Coder} and \textit{GPT-5.4}} with two strategies \ie{baseline prompting~\cite{DBLP:journals/corr/abs-2509-24419} and multi-stage optimization prompting from RQ3} under four context configurations \ie{hunk-\,+\,failure log, method-\,+\,failure log, file-\,+\,failure log, and file-level\,+\,failure log\,+\,top-5 support} across complexity levels, where complexity is measured by the test patch adaptation degree and the CUT change scale, respectively.\looseness=-1

\textbf{\textit{Higher repair scenario complexity generally leads to lower repair performance, while richer context mitigates this effect.}} 
As shown in Figure~\ref{fig:RQ5_test}, for all three LLMs, \textit{exec} declines across all prompting strategies and context configurations as the test patch adaptation degree increases. 
Taking \textit{DeepSeek-V3.2} as an example, the degradation is most notable when the number of changed code lines in the test patch exceeds 10, decreasing by 29.2\% on average (Figure~\ref{fig:RQ5_a1}). 
Regarding the complexity in terms of changed code lines in the CUT (Figure~\ref{fig:RQ5_c1}), six out of eight experimental groups peak at 4--10 changed lines and gradually degrade to their performance bottom beyond 100 changed lines. 
This trend indicates that when extensive modifications are made to the CUT, repairing becomes more challenging for LLMs. 
In Figure~\ref{fig:RQ5_d1}, as the number of changed methods in the CUT increases, method- and hunk-level context degrade in \textit{exec} while file-level context improves with a decreasing growth rate, suggesting that richer context mitigates degradation when the CUT changes extensively.
We can observe that other LLMs exhibit similar trends.

\textbf{\textit{Lower-level CUT change context exhibit advantages in providing focused information when the CUT change scale is small.}} 
As shown in Figures~\ref{fig:RQ5_c1}--\ref{fig:RQ5_c3}, when only 1--3 lines change in the CUT, hunk-level CUT context outperforms method-level context for all three LLMs with multi-stage prompting, and for \textit{DeepSeek-V3.2} it even surpasses file-level context by 16.6\% (Figure~\ref{fig:RQ5_c1}).
Similarly, when no more than 3 methods change in the CUT, hunk-level context outperforms method-level context by 10.1\% for \textit{DeepSeek-V3.2} with baseline prompting (Figure~\ref{fig:RQ5_d1}).
This might because hunk-level CUT changes offer more focused and precise information for test repair when the CUT change scale is small, exhibiting advantages in simpler co-evolution scenarios.
However, this advantage is model-dependent and diminishes when the CUT change scale is measured by the number of changed methods (Figures~\ref{fig:RQ5_d1}--\ref{fig:RQ5_d3}).

\begin{tcolorbox}[boxsep=2pt,left=2pt,right=2pt,top=2pt,bottom=2pt,width=\linewidth,colback=white!90!black,boxrule=0pt, colbacktitle=white!,toptitle=2pt,bottomtitle=1pt,opacitybacktitle=0]
\textbf{Answer to RQ4.}
Higher scenario complexity generally leads to lower repair success rates while richer context can mitigate this to some extent. 
In simple co-evolution scenarios, lower-level CUT changes provide more focused information and perform competitively. 
\end{tcolorbox}

\subsection{RQ5. Generalization to Agent-Based Test Repair}
In RQ5, we examine whether the helpful test repair context identified in RQ1 also benefits the autonomous LLM agents, which are equipped with powerful coding harnesses that orchestrate iterative LLM actions. 
We adopt \textit{Codex}~\cite{codex}, a SOTA coding agent that integrates well with GPT-5.4, which performs best in our experiments. 
Specifically, we compare a baseline run and a context-enhanced run of \textit{Codex}, both using the same Codex version (0.142.0-alpha.1), model (\textit{GPT-5.4} with \textit{high} thinking effort), and harness. 
The only difference is that the context-enhanced run is explicitly provided with the three contexts. 

To be more specific, for each repair, we provide the agent with two workspaces: the project at version $V$ and the project at version $V'$. 
In the $V'$ workspace, every test file modified by the commit, including the file containing the broken test, is reverted to its version-$V$ content, so that no developer-written test patch from the same commit is visible to the agent. 
Test utility classes and test resources changed by the commit are treated as part of the code under test and kept at version $V'$, consistent with the Cause context in RQ1. 
To keep dynamic validation possible, the other reverted test files are excluded from test compilation. We remove all Git-related information from both workspaces so that the agent cannot access the repository history or the ground-truth repair. 
The agent is only allowed to modify the specified test file, while retaining the tools needed for dynamic validation (including shell, Python, third-party libraries, and Maven), and is instructed to produce a single final repair. 

Given the time and budget cost of running the agent at scale, we apply it to a challenging subset of our benchmark to study whether the guidance derived from our empirical study can benefit the powerful agents. 
Specifically, we identify the 155 test repair tasks that \textit{GPT-5.4} failed under the best-performing context configuration \ie{Symptom\,+\,Cause at file-level\,+\,Top-5 Support} in RQ1. 
We further focus on cases involving large-scale CUT changes by selecting co-evolution cases introduced by commits modifying at least 100 Java source files, yielding 27 challenging cases from 6 projects. These cases represent the challenging end in our benchmark, where LLMs with only test repair instructions struggle.\looseness=-1  

The experimental results show that directly applying the agent \textit{Codex} to these 27 cases yields 12 (44.4\%) correct repairs, while providing the three types of context information improves the agent's performance to 20 (74.1\%) correct repairs \ie{+29.6 percentage points, 95\% CI: [+11.1, +48.1]~pp by paired bootstrapping over cases with 10,000 resamples}, corresponding to a 66.7\% relative improvement (95\% CI: [+16.7\%, +175.0\%]). 
This suggests that the guidance summarized from our study can also effectively enhance the test repair performance of powerful agents. 

To understand how the provided context helps the agent, we manually analyze the cases where it fails to repair the broken tests without context while successfully repairing them with context. 
We found that although the agent can search across an entire codebase and compare changes between two versions, it may not always correctly locate the critical information for test repair. 
We observe that providing clear Symptom, Cause, and Support contexts helps the agent in three important ways.
\begin{itemize}[leftmargin=*,topsep=1pt,itemsep=1pt]
     \item ($W1$) Context directs the agent to repair the correct API adaptation point instead of rewriting downstream assertions. 
     \item ($W2$) Context helps the agent preserve the original test intent, preventing the agent from drifting to an unrelated execution path without producing a passing patch.  
     \item ($W3$) The three types of context complement each other in enabling the agent to locate the actual relevant CUT change behind the observed failure behavior. 
\end{itemize}

We present two cases to illustrate how Codex benefits from these contexts. 
After source code changes made by commit \code{8805b3e3d6} \cite{iotdbcommit8805b3e3d6} in project \textit{apache/iotdb}, without the three types of context provided by us, \textit{Codex}'s repair attempt modified the result-checking logic ($W1$), incorrectly assuming that the spill-reader behavior had changed. 
The actual issue lies in the test-fixture code that no longer matched the new API \ie{\code{DiskSpiller} had become abstract and \code{SortBufferManager} now required constructor arguments}. 
In this case, the Symptom context exposed the exact compilation mismatch error, the Cause context localized the change to \code{DiskSpiller} and \code{SortBufferManager} and the Support context exposed their new API shape ($W3$). 
With these cues, \textit{Codex} recognized this API mismatch and repaired the fixture setup directly while keeping the original assertions unchanged ($W1$). 

Another example is about the broken test caused by commit \code{cda0888352} \cite{pinpointcommitcda0888352} in project \textit{pinpoint-apm/pinpoint}. 
Without context provided, Codex bypassed the original \code{PluginTestAgent} bootstrap path and rewrote the test to instantiate \code{DefaultApplicationContext}, changing the program behavior that was intended to be validated by the original test while still not leading to a successful repair ($W2$). 
In this case, the Symptom context revealed the incompatibility around \code{AgentOption}, the Cause context localized the relevant CUT change to \code{PluginTestAgent} and \code{OverrideModuleFactory}, and the Support context exposed the updated constructor shape of \code{PluginTestAgent} ($W3$). 
With such guidance, \textit{Codex} preserved the original bootstrap scenario and repaired it by supplying the missing \code{agentPath} ($W2$), which was the minimal change needed to satisfy the updated startup path.

\vspace{-1mm}
\begin{tcolorbox}[boxsep=2pt,left=2pt,right=2pt,top=2pt,bottom=2pt,width=\linewidth,colback=white!90!black,boxrule=0pt, colbacktitle=white!,toptitle=2pt,bottomtitle=1pt,opacitybacktitle=0]
\textbf{Answer to RQ5.}
Providing the context that proves helpful for LLMs in previous RQs effectively improves Codex's performance in test repair by 66.7\% on challenging cases. 
Case studies show that such contexts help the agent choose the correct repair target, avoid drifting away from the original behavior, and locate the actual relevant CUT change. 
\end{tcolorbox}

\section{Threats to Validity}\label{sec:6}

\textbf{Internal validity.}
LLM outputs can vary with temperature~\cite{DBLP:journals/tosem/OuyangZHW25}. 
To observe representative model behaviors and obtain meaningful conclusions, we set the temperature to 0.6 per vendor recommendations. 
Besides, our multi-attempt repair experiments in RQ2 show that each additional independent attempt yields a steady, consistent performance gain, confirming that individual results are not artifacts of random variance. 
Second, the quality of Support context may affect repair performance. To ensure its quality, we employ a frontier LLM, \textit{DeepSeek-V3.2}, to select the relevant unchanged methods and classes. 
To ensure a fair comparison, we supply identical Support context to all three LLMs. 
Third, flaky tests and our unified build environment (Java 17, Maven 3.9.9) may cause spurious test failures or passes. To mitigate this, we executed each instance three times during benchmark construction and retained only those with consistent outcomes. 

\textbf{External validity.}
Our study subjects target JUnit tests in Java projects managed by Maven, which may not represent other ecosystem settings. 
But Maven and JUnit are among the most widely adopted Java toolchains, and the co-evolution patterns and investigation dimensions we study are not inherently language-specific. 
We also evaluate three LLMs from diverse providers to reduce model-specific bias. 
Moreover, the reasoning action taxonomy is derived from the traces of \textit{DeepSeek-V3.2}, and the effectiveness of context types is evaluated on our benchmark; these findings may not fully generalize to other models or benchmarks. 
We partially mitigate this by applying the taxonomy-derived strategies to all three LLMs (RQ3) and transferring the context findings to the \textit{Codex} agent (RQ5). 
Finally, our benchmark covers commits from May~2024 to April~2026, partially post-dating the training cutoffs of modern LLMs, which reduces but does not eliminate the risk of data contamination. 

\textbf{Construct validity.}
Test pass rate (\textit{exec}) does not guarantee semantic equivalence with developer intent, as a repaired test may pass by weakening assertions. 
This metric is nonetheless standard in prior test repair work~\cite{DBLP:journals/corr/abs-2509-24419, DBLP:conf/issre/LiuYX0024}, and we mitigate this by verifying repair quality against developer-committed patches via both LLM-as-a-Judge and human inspection. 
The LLM judge and independent human auditor reach strong agreement that only very few cases weakened the test oracle on the sampled repairs (Section~\ref{sec:3}).
Nevertheless, the judge only detects degenerate oracle weakening \eg{assertion removal} rather than verifying full semantic correctness, so subtler deviations from developer intent may remain undetected.

\section{Related Work}\label{sec:5}

\textbf{Empirical Studies on Source--Test Co-evolution.}
A long line of empirical work has characterized how production and test code co-evolve.
Early repository-mining studies examined whether and when test code changes together with production code, using co-change analysis~\cite{DBLP:conf/icst/ZaidmanRDD08}, association rules~\cite{DBLP:conf/msr/LubsenZP09}, and fine-grained change patterns~\cite{DBLP:conf/scam/MarsavinaRZ14, DBLP:conf/icsm/LevinY17}.
Pinto et al.~\cite{DBLP:conf/sigsoft/PintoSO12} studied real-world test-suite evolution and found that repairing broken tests is only one driver of test changes, alongside the more frequent test additions, deletions, and refactorings.
Subsequent studies characterized code-to-test co-evolution patterns~\cite{DBLP:conf/wcre/Wang000W21, DBLP:conf/icst/ShimmiR22} and leveraged them to identify obsolete tests~\cite{DBLP:conf/wcre/Wang000W21} or recommend test updates~\cite{DBLP:conf/ast/ShimmiR22}.
Sun et al.~\cite{DBLP:journals/tosem/SunYLXLL23} revisited how co-evolution samples are mined, revealing that the widely adopted co-change assumption introduces pervasive noise, and proposed CHOSEN to identify genuine production--test co-evolution samples; Huang et al.~\cite{DBLP:journals/stvr/HuangTCZ24} similarly predict whether test code should co-change with production code.
Guo et al.~\cite{DBLP:journals/tse/GuoCXCLZ25} studied the technical debt incurred by unbalanced production--test co-evolution in 15 open-source projects and proposed an approach to identify and quantify it.
Closer to the use of LLMs, Lemner et al.~\cite{DBLP:journals/corr/abs-2409-06416} explored which industrial test maintenance activities LLMs can support.
These studies focus on understanding \textit{whether}, \textit{when}, and \textit{why} tests co-evolve with production code, or on detecting tests that need to be updated, but do not investigate \textit{how} broken tests should be repaired.
Meanwhile, existing evaluations of LLM-based test repair are conducted as part of proposing new techniques, each fixing a particular choice of repair context and pipeline.
Very recent studies have begun to evaluate LLMs for test evolution, yet they either examine prompt design under single-focal-method changes~\cite{10.1145/3832282} or focus on the overall performance of coding agents~\cite{DBLP:journals/corr/abs-2605-06125, DBLP:journals/corr/abs-2607-02469}.
Different from existing work, our work is the first to systematically and empirically investigate the design space of LLM-based test repair, \ie{repair context, pipeline orchestration, explicit reasoning, scenario complexity, and generalization to coding agents}, on realistic co-evolution scenarios involving multiple changed methods and classes.

\textbf{Test Repair.}
Early rule-driven techniques~\cite{DBLP:conf/kbse/DanielJDM09, DBLP:conf/icsm/MirzaAghaeiPP10, DBLP:conf/issta/DanielGM10, DBLP:conf/apsec/XuHWY14, DBLP:conf/sigsoft/MirzaAghaei11, DBLP:conf/icst/MirzaAghaeiPP12, DBLP:journals/stvr/MirzaAghaeiPP14, DBLP:conf/icst/LidO19} fixed obsolete test cases using handcrafted rules or predefined patterns, limiting their applicability to diverse repair scenarios. 
Deep learning later enabled learned repair models such as CEPROT~\cite{DBLP:conf/kbse/HuLXLXY23} and TaRGET~\cite{DBLP:journals/tse/YaraghiHKB25}. 
More recently, LLM-based approaches have emerged: SYNTER~\cite{DBLP:conf/issre/LiuYX0024} was the first to address test failures caused by changes in focal method signatures. REACCEPT~\cite{DBLP:journals/pacmse/ChiWHYCSS25} leverages dynamic validation feedback to improve repair performance. 
UTFix~\cite{DBLP:journals/pacmpl/RahmanKCGWMDR25} prompts LLMs with static and dynamic code slices to repair tests and TestUpdater~\cite{DBLP:journals/corr/abs-2509-24419} uses LLMs to both repair and enhance test cases in response to focal method modifications. 
MuMuTestUp~\cite{DBLP:journals/pacmse/TianLPZCSS26} is a very recent multi-agent technique that updates tests with mutation-guided assertion refinement. 
Among these techniques, TaRGET focuses on single-hunk test patch generation and the others repair obsolete tests caused by a single focal method change. 
MuMuTestUp relaxes this assumption by also considering changes along the call chain, but still centers its analysis on a single focal method. 
Different from existing work, our work addresses more complex and practical test repair scenarios. 

\textbf{Test Generation.} 
Automated test generation reduces the manual burden of constructing test cases. 
Conventional tools like EvoSuite~\cite{DBLP:conf/sigsoft/FraserA11} and Symstra~\cite{DBLP:conf/tacas/XieMSN05} are based on evolutionary strategies, random testing, or symbolic execution, while learning-based approaches such as AthenaTest~\cite{tufano2021unittestcasegeneration} (built on BART~\cite{DBLP:conf/acl/LewisLGGMLSZ20}), Atlas~\cite{DBLP:conf/icse/WatsonTMBP20} and Tufano et al.~\cite{DBLP:conf/ast/TufanoDSS22} target test or assertion generation, though often at the cost of syntactic validity or executability~\cite{DBLP:journals/csur/YangXLG22}. 
Recent LLM-based approaches, including ChatTester~\cite{DBLP:journals/pacmse/Yuan0DW00L24} which is the first LLM-based generator, ChatUniTest~\cite{DBLP:conf/sigsoft/ChenHZHDY24} adopting a generate–verify–repair workflow to enhance test accuracy and coverage, HITS~\cite{DBLP:conf/kbse/WangL0J24} that employs code slicing to decompose complex methods into simpler components. 
Unlike these methods, which generate new tests for individual methods, our work focuses on repairing and enhancing existing tests during software evolution. 

\section{Conclusion}\label{sec:7}

This paper presented the first systematic empirical study of LLM-based test repair for source--test co-evolution, grounded in a new benchmark of 500 dynamically validated, real-world instances from 15 highly-starred open-source Java projects. 
Across five research questions, we investigated key design dimensions including repair context, pipeline orchestration, LLM reasoning actions, repair scenario complexity, and agent-based generalization. 
Our study reveals that test failure symptoms, source code changes, and supportive code context complementarily contribute to effective test repair, and that iterative self-repair rounds further benefit LLMs. 
We developed a reasoning action taxonomy from LLM reasoning traces on successful repairs. 
It helps incentivize cost-effective LLM-based test repairs. 
We also demonstrate that our empirically derived insights translate into tangible improvements for coding agents. 
Our findings provide empirical evidence and offer actionable guidelines for effective LLM-based test repair.

{
\section{Data Availability}
\label{sec:8}
The replication package is available at \url{https://anonymous.4open.science/r/Source-Test-Co-evolution-Study-F4F4}. 
}

% References
\bibliographystyle{ACM-Reference-Format}
\bibliography{bibliography/references,bibliography/links}

\end{document}